\documentclass[
  aps,
  preprintnumbers,
  prd,
  10pt,
  twocolumn,
  floatfix,
  nofootinbib,
  superscriptaddress
]{revtex4-2}

\usepackage{graphicx} 
\usepackage{bm}       
\usepackage{multirow}
\usepackage{xcolor}
\usepackage{subfigure}
\usepackage{adjustbox}
\usepackage{anyfontsize}
\usepackage{amsfonts} 
\usepackage{amsmath}  
\usepackage{amsthm}   
\usepackage{amssymb}  
\usepackage{mathrsfs}
\usepackage{bbold}
\usepackage{placeins}
\usepackage{mathtools}
\usepackage{enumitem}
\usepackage{calc}
\usepackage[
  unicode=true,
  bookmarks=false,
  breaklinks=false,
  pdfborder={0 0 1},
  backref=false,
  colorlinks=true
]{hyperref}
\hypersetup{
  citecolor=blue, 
  linkcolor=blue, 
  urlcolor=blue
}
\usepackage{orcidlink} 
\usepackage[normalem]{ulem}

\graphicspath{{./figs/}}

\allowdisplaybreaks

\newcommand{\Tr}{\text{Tr}}

\makeatletter
\newcommand{\ecitem}[2]{%
  \item[\textbf{#1}]%
  \phantomsection
  \protected@edef\@currentlabel{\unexpanded{\textbf{#1}}}%
  \label{#2}%
}
\makeatother

\newcommand{\orcidauthorOHNO}{0000-0003-1798-8222}
\newcommand{\orcidauthorTOMIYA}{0000-0001-9374-3716}
\newcommand{\orcidauthorCHOI}{0000-0002-5438-5490}

\begin{document}
\preprint{UTHEP-826, UTCCS-P-182}

\title{Bias-Corrected Machine-Learning Estimation of Chiral Condensate
  Cumulants:\\A Retrospective Lattice QCD Case Study}

\author{Benjamin J. Choi\,\orcidlink{\orcidauthorCHOI}}
\email{benchoi@het.ph.tsukuba.ac.jp}
\affiliation{Center for Computational Sciences, University of Tsukuba,
  1-1-1 Tennodai, Tsukuba, Ibaraki 305-8577, Japan}

\author{Hiroshi Ohno\,\orcidlink{\orcidauthorOHNO}}
\email{hohno@ccs.tsukuba.ac.jp}
\affiliation{Center for Computational Sciences, University of Tsukuba,
  1-1-1 Tennodai, Tsukuba, Ibaraki 305-8577, Japan}

\author{Akio Tomiya\,\orcidlink{\orcidauthorTOMIYA}}
\email{akio@yukawa.kyoto-u.ac.jp}
\affiliation{Department of Information and Mathematical Sciences,
  Tokyo Woman's Christian University, 2-6-1 Zempukuji, Suginami-ku,
  Tokyo 167-8585, Japan}
\affiliation{RIKEN Center for Computational Science,
  7-1-26 Minatojima-minami-machi, Chuo-ku, Kobe 650-0047, Japan}
\affiliation{Department of Physics, Kyoto University, Kitashirakawa,
  Sakyo-ku, Kyoto 606-8502, Japan}

\date{\today}

\begin{abstract}
  We present a retrospective case study of bias-corrected machine
  learning (ML) estimates of traces of the inverse Dirac operator,
  $\Tr\,M^{-n}$ ($n=1,2,3,4$), using a fixed lattice QCD dataset and
  examining how the results depend on the relative proportions of the
  labeled and training sets.
  Two supervised learning approaches are examined: one using $\Tr \,
  M^{-1}$ as the input feature, and the other employing gauge
  observables such as the plaquette and rectangle.
  Beyond the direct estimation of $\Tr \, M^{-n}$, we further
  investigate two derived applications of the ML estimations: the
  evaluation of the cumulants of the chiral condensate within a single
  ensemble and that obtained through multi-ensemble reweighting across
  ensembles with different quark masses.
  Within this fixed dataset, the bias-corrected estimates show close
  agreement with the full-data reference under the adopted evaluation
  criteria, while the uncorrected estimates can exhibit amplified
  deviations after the nonlinear cumulant and reweighting steps.
  For the approach using $\Tr\,M^{-1}$ as the input feature, nominal
  solve-count accounting suggests that the Dirac-inversion cost could
  be reduced to approximately $25.75\%$ of that of the conventional
  calculation in the present setup.
  This value is a cost projection rather than an end-to-end benchmark:
  it assumes comparable costs for successive inversions and excludes
  model-training and analysis overhead.
\end{abstract}

\maketitle

\section{Introduction}
\label{sec:intro}

Chiral symmetry plays a central role in determining the
finite-temperature phase structure of Quantum Chromodynamics (QCD)
with light quarks.
In the massless limit, the QCD Lagrangian possesses an
$\mathrm{SU}(N_f)_L \times \mathrm{SU}(N_f)_R$ symmetry, which is
spontaneously broken at low temperatures, leading to a nonvanishing
chiral condensate and associated Nambu–Goldstone modes.
At high temperatures, this symmetry is restored through a phase
transition or crossover, depending on the quark masses.
The quark mass dependence of the thermal transition is summarized in
the Columbia plot \cite{Brown:1990ev}, which maps the transition order
in the plane of light and strange quark masses.
In the massless limit, symmetry-based arguments suggest a second-order
transition for two flavors under the corresponding assumptions and a
first-order transition for three flavors \cite{Pisarski:1983ms}, while
for physical quark masses, lattice QCD indicates a crossover at
vanishing baryon chemical potential.
These regions are separated by lines of second-order critical points
belonging to the $Z(2)$ universality class.
At finite baryon chemical potential, the crossover may turn into a
first-order transition via a critical endpoint; whether and where such
an endpoint occurs remains a central open problem in QCD
thermodynamics.
For more details, see \cite{Philipsen:2021qji, Guenther:2020jwe} and
references therein.

Within the framework of lattice QCD, locating the critical point
requires the computation of higher-order cumulants of the chiral
condensate, which serves as the order parameter.
In particular, fourth-order cumulants are sensitive to the underlying
universality class and form the basis of the kurtosis intersection
method, which is widely used to determine the critical point
\cite{Jin:2014hea, Kuramashi:2016kpb}.
The evaluation of these cumulants requires stochastic estimates of
traces of inverse powers of the Dirac operator, $\textrm{Tr}\, M^{-n}$
with $n =1,2,3,4$.
This, in turn, involves repeatedly solving systems of linear equations
associated with the Dirac matrix.
We note that similar high-order cumulants are also studied for
conserved charges, such as baryon number, electric charge, and
strangeness, whose fluctuations provide complementary probes of the
QCD phase structure in both lattice calculations and heavy-ion
collision experiments \cite{Stephanov:2008qz, Bazavov:2020bjn,
  Borsanyi:2018grb}.
Their evaluation involves computational requirements closely analogous
to those addressed here.
Since the Dirac operator is an extremely large sparse matrix, these
linear systems are typically solved using iterative Krylov methods.
Even with stochastic trace estimators such as the Hutchinson method
\cite{Dong:1993pk}, the computational cost remains substantial.

Ref.~\cite{Yoon:2018krb} adopts a regression-based approach based on
machine learning (ML) combined with a bias correction method that
applies the idea of All Mode Averaging method \cite{Bali:2009hu,
  Blum:2012uh} to reduce, as much as possible, the prediction bias
inevitably present in ML estimations.
This bias correction method is essentially based on supervised
learning.
Accordingly, the entire dataset is divided into a labeled set and an
unlabeled set.
Furthermore, the labeled set is partitioned into a training set and a
bias correction set.
Because the labeled set equals the union of the training and bias
correction sets, varying the training-set fraction induces a
corresponding trade-off adjustment in the bias correction fraction.

In the present retrospective case study, we adopt this approach and
systematically vary the relative proportions of the labeled set and,
within it, the training set.
We then perform bias-corrected ML estimations of $\Tr \, M^{-n}$ for
various ratios of the labeled and training sets, and examine how the
agreement with the full-data reference changes with these ratios.
Beyond this analysis, we further carry out two practical applications
derived from the ML estimations.
First, we combine the ML estimation results on $\Tr \, M^{-n}$ to
compute the cumulants within a single ensemble and assess how closely
these ML-derived quantities agree with the full-data results.
Second, we extend the same idea to multiple ensembles generated at
different quark masses by performing multi-ensemble reweighting, and
evaluate the agreement of the resulting cumulant estimates with the
full-data results at the reweighted target points.
The purpose of these applications is to further explore whether the ML
estimations can be effectively utilized in practical physics analyses
by examining how well the ML-derived quantities constructed from these
estimations reproduce the physical observables of interest.

In addition to our main results obtained with bias correction, we also
perform parallel calculations without applying any bias correction, in
which the entire labeled set is used for model training without
reserving a separate bias correction set.
This allows us to investigate how the absence of bias correction
affects the resulting estimations.
For validation purposes, our ML estimations are not limited to the
direct prediction of $\Tr \, M^{-n}$, but also extend to the
ML-derived quantities such as cumulants computed within a single
ensemble and those obtained through multi-ensemble reweighting.
Throughout these analyses, we perform the conventional calculations
simultaneously, using the original (non-split) datasets without any
data partitioning, and compare them with the ML estimations to assess
the resulting agreement.
All comparisons reported here are conditional on the fixed dataset and
the deterministic partition used in this study; they are intended as a
retrospective assessment of feasibility rather than a demonstration of
dataset-independent performance.

This paper is organized as follows.
In Sec.~\ref{sec:formalism} we summarize the notation and conventions
frequently used throughout this work.
We also define the cumulants required to apply the ML estimations on
the traces for the evaluation of the chiral condensate cumulants.
In Sec.~\ref{sec:framework} we describe the methodological framework
adopted in this study, including the procedure for computing the ML
estimations and for splitting the data.
In Sec.~\ref{sec:trace} we present the analysis of the estimation
agreement for the ML estimates of the trace of the inverse Dirac
operator, $\Tr\,M^{-n}$, evaluated for various ratios of the labeled
and training sets.
In Sec.~\ref{sec:cumulant-single} we present the first application of
the derived ML estimations, namely the cumulant evaluation within a
single ensemble, and discuss agreement as a function of the labeled
and training set ratios.
In Sec.~\ref{sec:cumulant-reweighting} we describe the second
application of the derived ML estimations, which concerns the cumulant
evaluation through multi-ensemble reweighting across ensembles with
different quark masses, and analyze the agreement of the corresponding
ML estimation results with the full-data reference.
In Sec.~\ref{sec:conc} we conclude.

\section{Formalism and Dataset}
\label{sec:formalism}

Before presenting the details of the analysis procedure, we summarize
the notation and conventions used in this work in
Table~\ref{tab:notation-1} for clarity and reference throughout the
following sections.

\begin{table}[tb]
  \renewcommand{\arraystretch}{1.0}
  \caption{Notation and convention used in this paper for the
    explanation of our work.
    Here, $Z$ collectively denotes $(X,Y)$, and all dataset sizes
    satisfy $\lvert S^X_\ast\rvert=\lvert S^Y_\ast\rvert$.}
  \label{tab:notation-1}
  \begin{ruledtabular}
    \begin{tabular}{@{\quad}r@{\quad}|@{\quad}p{0.69\columnwidth}@{\quad}}
    Symbol & Description \\
    \hline
    $X$ & input observables (or feature) \newline (\textit{e.g.},~$X =
    \text{Plaquette}$, $\text{Rectangle}$, $\Tr \, M^{-1}$) \\[0.5em]
    $Y$ & output observables (or target) \newline (\textit{e.g.},~$Y =
    \Tr \, M^{-n}$ for $n=1,2,3,4$) \\[0.5em]
    $S^Z$ & the total dataset of $Z=X,Y$ \newline where $S^Z =
    S^Z_{\text{LB}} \cup S^Z_{\text{UL}}$ \\[0.5em]
    $S^Z_{\text{LB}}$ & the labeled set of the original data \newline
    where $S^Z_{\text{LB}} = S^Z_{\text{TR}} \cup
    S^Z_{\text{BC}}$ \\[0.5em]
    $S^Z_{\text{TR}}$ & the training set of the original data \\[0.5em]
    $S^Z_{\text{BC}}$ & the bias correction set of the original
    data \\[0.5em]
    $S^Z_{\text{UL}}$ & the unlabeled set of the original
    data \\[0.5em]
    $N$ & the number of elements of $S^Z$ \newline where $N =
    \left|S^{X}\right| = \left|S^{Y}\right|$ \\[0.5em]
    $N_{\text{LB}}$ & the number of elements of
    $S^Z_{\text{LB}}$
    \\[0.5em]
    $N_{\text{TR}}$ & the number of elements of
    $S^Z_{\text{TR}}$
    \\[0.5em]
    $N_{\text{BC}}$ & the number of elements of
    $S^Z_{\text{BC}}$
    \\[0.5em]
    $N_{\text{UL}}$ & the number of elements of
    $S^Z_{\text{UL}}$
    \\[0.5em]
    $f(X)$ & the model trained with $S^{X}_{\text{TR}}$ and
    $S^{Y}_{\text{TR}}$ \\[0.5em]
    $Y^P$ & the ML estimation on $Y$ \\[0.5em]
    $S^P_{\text{BC}}$ & the bias correction set composed of
    the \newline ML estimations $Y^P_{\text{BC}}$ corresponding to
    $S^Y_{\text{BC}}$ \\[0.5em]
    $S^P_{\text{UL}}$ & the unlabeled set composed of the \newline ML
    estimations $Y^P_{\text{UL}}$ corresponding to $S^Y_{\text{UL}}$
    \\
    \end{tabular}
  \end{ruledtabular}
\end{table}

\begin{table}[tb]
  \renewcommand{\arraystretch}{1.2}
  \caption{Data used in this paper, originally produced for
    Ref.~\cite{Ohno:2018gcx}.
    The gauge coupling and the clover coefficient are fixed to $\beta
    = 1.60$ and $c_{\text{SW}}=2.065$ respectively, for all five
    ensembles.
    Here, $N$ follows the same convention as in
    Table~\ref{tab:notation-1}, where it corresponds to the number of
    configurations.}
  \label{tab:ensemble-1}
  \begin{ruledtabular}
    \begin{tabular}{@{\quad}c@{\quad}|@{\quad}c@{\quad}|@{\quad}c@{\quad}|@{\quad}c@{\quad}}
    \texttt{ID} & $N_{\text{S}}^3 \times N_{\text{T}}$ & $\kappa$ & $N$ \\
    \hline
    \texttt{L12T4b1.60k13575} & $12^3 \times 4$ & 0.13575 & 20000 \\
    \texttt{L12T4b1.60k13577} & $12^3 \times 4$ & 0.13577 & 20000 \\
    \texttt{L12T4b1.60k13580} & $12^3 \times 4$ & 0.13580 & 20000 \\
    \texttt{L12T4b1.60k13582} & $12^3 \times 4$ & 0.13582 & 20000 \\
    \texttt{L12T4b1.60k13585} & $12^3 \times 4$ & 0.13585 & 20000 \\
    \end{tabular}
  \end{ruledtabular}
\end{table}

To evaluate the ML estimation framework retrospectively in a lattice
QCD setting, we employ an existing dataset originally generated for
Ref.~\cite{Ohno:2018gcx}.
These configurations were produced on the \texttt{Oakforest-PACS}
system~\cite{Boku:2017urp} using the \texttt{BQCD}
code~\cite{Nakamura:2010qh} with the $N_{\text{f}}=4$ Wilson Clover
quark action~\cite{Sheikholeslami:1985ij} and the Iwasaki gauge
action~\cite{Iwasaki:1985we, Iwasaki:1983iya}.
The same dataset has been used in previous studies of the chiral phase
transition and provides the fixed case-study data for the present
bias-corrected ML analysis.
The ensembles used in this analysis are listed in
Table~\ref{tab:ensemble-1}.
The ensembles primarily examined in this work correspond to the
lattice volume $V = 12^3 \times 4$ and coupling $\beta = 1.60$,
comprising five distinct $\kappa$ values.
In Ref.~\cite{Ohno:2018gcx}, measurements were performed for several
$\beta$ values ($1.60$, $1.61$, and $1.62$) on the same volume in
order to determine the critical endpoint parameters $\beta_E$ and
$K_E$ via the kurtosis intersection method.
Among these, the ensembles at $\beta = 1.60$ are found to lie in the
first-order phase transition region, whereas those at $\beta = 1.62$
appear to belong to the crossover regime.

The quark operator employed in this study is the Wilson-Clover
operator, which is defined as
\begin{align} 
  M(x,y) &= \frac{1}{2\kappa} \, \delta_{x,y} + \frac{ \mathrm{i} }{4}
  \, c_\textrm{sw} \, \sigma_{\mu\nu} F_{\mu\nu}(x) \, \delta_{x,y}
  \nonumber \\
  &\hphantom{=\ } - \frac{1}{2} \sum_{\mu=1}^{4}\sum_{s=\pm 1} \left(
  1 - s \,\gamma_\mu \right) \, U_{s\,\mu}(x) \, \delta_{x,\,y +
    s\,\hat\mu} \,.
\end{align}
Building on the operator introduced above, we define the quark-loop
contributions (moment operators of the chiral condensate) as
\begin{align}
  Q_1 &= N_\textrm{f} \; \Tr \, M^{-1} \,, \label{eq:Q_1} \\
  Q_2 &= - \, \left( N_\textrm{f} \; \Tr \, M^{-2} \right) + \left(
  N_\textrm{f} \; \Tr \, M^{-1} \right)^2 \,, \label{eq:Q_2} \\
  Q_3 &= 2 \, \left( N_\textrm{f} \; \Tr \, M^{-3} \right) - 3 \,
  \left( N_\textrm{f} \; \Tr \, M^{-2} \right) \left( N_\textrm{f} \;
  \Tr \, M^{-1} \right) \nonumber \\
  &\hphantom{=\ } + \left( N_\textrm{f} \; \Tr \, M^{-1} \right)^3
  \,, \label{eq:Q_3} \\
  Q_4 &= -6 \, \left( N_\textrm{f} \; \Tr \, M^{-4} \right) + 8 \,
  \left( N_\textrm{f} \; \Tr \, M^{-3} \right) \left( N_\textrm{f} \;
  \Tr \, M^{-1} \right) \nonumber \\
  &\hphantom{=\ } + 3\big(N_\textrm{f}\;\Tr \, M^{-2}\big)^2 - 6 \,
  \left( N_\textrm{f} \; \Tr \, M^{-2} \right) \left(N_\textrm{f}\;\Tr
  \, M^{-1} \right)^2 \nonumber \\
  &\hphantom{=\ } + \left(N_\textrm{f}\;\Tr \, M^{-1}\right)^4
  \,. \label{eq:Q_4}
\end{align}
For completeness, the derivations of
Eqs.~\eqref{eq:Q_1}--\eqref{eq:Q_4} are given in
Appendix~\ref{sec:cumulants-app}.

The chiral condensate cumulants up to the fourth order are defined as
\begin{align}
  \Sigma &= \frac{C_{1}}{V} \,, \label{eq:cond} \\
  \chi &= \frac{C_{2}}{V} \,, \label{eq:susp} \\
  S &= \frac{C_{3}}{C_{2}^{\frac{3}{2}}} \,, \label{eq:skew} \\
  K &= \frac{C_4}{C_{2}^{2}} \,, \label{eq:kurt}
\end{align}
where $V = N_{\text{S}}^3 \times N_{\text{T}}$ denotes the lattice
volume.
The corresponding $C_i$ is written in terms of $\langle Q_i \rangle$
as
\begin{align}
  C_{1} &= \left\langle Q_1 \right\rangle \,, \label{eq:C_1}
  \\[0.25em]
  C_{2} &= \left\langle Q_2 \right\rangle - \left\langle Q_1
  \right\rangle^{2} \,, \label{eq:C_2} \\[0.25em]
  C_{3} &= \left\langle Q_3 \right\rangle-3\left\langle Q_2
  \right\rangle \left\langle Q_1 \right\rangle + 2 \left\langle Q_1
  \right\rangle^{3} \,, \label{eq:C_3} \\[0.25em]
  C_{4} &= \left\langle Q_4 \right\rangle - 4 \left\langle Q_3
  \right\rangle \left\langle Q_1 \right\rangle -3 \left\langle Q_2
  \right\rangle^{2} \nonumber \\
  &\hphantom{=} +12 \left\langle Q_2 \right\rangle \left\langle Q_1
  \right\rangle^{2}-6\left\langle Q_1 \right\rangle^{4}
  \,. \label{eq:C_4}
\end{align}
Here, $\Sigma$, $\chi$, $S$, and $K$ denote the chiral condensate, its
susceptibility, skewness, and kurtosis, respectively.

Strictly speaking, $C_j$ and $\{\Sigma,\chi,S,K\}$ are distinct;
nevertheless, following common usage, we refer to these quantities
collectively as ``cumulants'' in what follows.
A detailed derivation of Eqs.~\eqref{eq:cond}--\eqref{eq:C_4} is
provided in Appendix~\ref{sec:cumulants-app}.

\section{Framework for ML Estimation}
\label{sec:framework}

\subsection{Supervised Learning with Bias Correction}

For the supervised ML estimation, we perform the following two steps:
\begin{enumerate}
\item We train a model $f(X)$ using the data in the $S^X_{\text{TR}}$
  and $S^Y_{\text{TR}}$.
\item We input $X \in S^X_{\text{UL}}$ into the model $f(X)$ and
  obtain ML estimation $f(X) = Y^P \approx Y$.
\end{enumerate}
Note that only $Z \in S^Z_{\text{LB}} \subset S^Z$ ($Z=X,Y$) can be
used for the model training, which means that there may exist various
unwanted fluctuations or unusual biases due to the \emph{partial} data
usage, in principle.

Therefore, using the methodology given in Ref.~\cite{Yoon:2018krb}, we
do not use all $S^Z_{\text{LB}}$ for model training but use only
$S^Z_{\text{TR}} \subset S^Z_{\text{LB}}$, the training set.
We use the remaining $S^Z_{\text{BC}} = S^Z_{\text{LB}} \setminus
S^Z_{\text{TR}}$ for the bias correction.

We then compute
\begin{align}
  \bar{Y}_{\mathcal{P}1} & = \frac{1}{N_{\textrm{UL}}} \sum_{Y_i \in
    S^Y_{\text{UL}}} Y_i^{P} + \frac{1}{N_{\textrm{BC}}} \sum_{Y_j \in
    S^Y_{\text{BC}}} \left( Y_j - Y_j^{P}\right) \,. \label{eq:P1-1}
\end{align}
An alternative weighted estimator, denoted $\mathcal{P}2$, was also
examined.  Because its uncertainty behavior requires additional care,
we use $\mathcal{P}1$ as the sole primary estimator and present the
definition and diagnostic results for $\mathcal{P}2$ in
Appendix~\ref{sec:app-p2}.

\subsection{Gradient boosting decision tree}
\label{sec:gbdt}

For the ML estimation of $Y = \Tr \, M^{-n}$, we use the gradient
boosting decision tree regression method \cite{Mason:1999zz,
  Friedman:2001zz}.
To be specific, we use the \texttt{LightGBM} \cite{Ke:2017zz}
framework via \texttt{JuliaAI/MLJ.jl} \cite{Blaom:2020zz}.
We use depth-3 trees with a learning rate of 0.1, a subsampling rate
of 0.7, and 40 boosting stages.
The tree depth, learning rate, and subsampling rate are chosen
following the setup of Ref.~\cite{Yoon:2018krb}, while the number of
boosting stages is set to 40 because preliminary tests suggested that
the validation loss had approximately reached a plateau around this
point.

\subsection{Method for Partitioning the Total Dataset $S^Z$}
\label{sec:dataset-split}

The total dataset $S^Z$ is partitioned deterministically into several
subsets used for supervised learning and bias correction.
Specifically, the full set of configuration indices $\{1,2,\dots,N\}$
is first divided into a labeled set $S^Z_{\text{LB}}$ and an unlabeled
set $S^Z_{\text{UL}}$.
The labeled set is then further split into a training set
$S^Z_{\text{TR}}$ and a bias-correction set $S^Z_{\text{BC}}$.

The partitioning procedure is fully deterministic and is designed so
that each subset samples the total dataset $S^Z$ as uniformly as
possible while preserving the original ordering of configurations.
All numerical results presented in this work are obtained using a
specific instance of this general partitioning scheme.
The detailed construction algorithm and its general formulation are
described in Appendix~\ref{sec:app-dataset-split}.
Consequently, the numerical comparisons in this retrospective study
are conditional on this single deterministic partition and do not
include split-to-split variability.

\subsection{Scan over labeled and training fractions}

For the economic efficiency of the ML estimation, that is, to reduce
the cost of the Dirac-operator inversions, we should use the least
possible $S^Z_{\text{LB}}$ ($Z=X,Y$).
Here, note that we need to use $Y$ as well as $Y^P$ for the bias
correction.
This means that we should grant sufficient statistics to the
$S^Z_{\text{BC}}$ among the $S^Z_{\text{LB}}$.
In summary, we monitor the parameters $\mathcal{R}_{\text{LB}}$ and
$\mathcal{R}_{\text{TR}}$, defined as
\begin{equation}
  \mathcal{R}_{\text{LB}} \equiv \frac{N_{\text{LB}}}{N}
  \,, \label{eq:R-LB-1}
\end{equation}
and
\begin{equation}
  \mathcal{R}_{\text{TR}} \equiv \frac{N_{\text{TR}}}{N_{\text{LB}}}
  \,. \label{eq:R-TR-1}
\end{equation}
Here, $N = N_{\text{LB}} + N_{\text{UL}}$ and $N_{\text{LB}} =
N_{\text{TR}} + N_{\text{BC}}$.  We scan $\mathcal{R}_{\text{LB}} =
1,\,2,\,\cdots,\,25\%$ and $\mathcal{R}_{\text{TR}} =
0,\,10,\,\cdots,\,100\%$ as representative fractions for this
retrospective analysis.
Here, we examine the case $\mathcal{R}_{\text{TR}} = 0 \%$
($S_{\text{TR}}^Z = \emptyset$ for $Z=X,Y$) to check that
$S^Y_{\text{LB}}$ itself approaches the true answer as
$\mathcal{R}_{\text{LB}}$ increases.
In this case, we do not perform ML estimation, but only observe the
statistical average and error of $S^Y_{\text{LB}}$.
We also examine the case $\mathcal{R}_{\text{TR}} = 100 \%$
($S_{\text{BC}}^Z = \emptyset$) to see what happens when the bias
correction method is not used in the ML estimation.

\subsection{Correlation Structure among Observables}

\begin{figure}[b!]
  \subfigure[$\kappa = 0.13575$, \texttt{L12T4b1.60k13575} (the
    heaviest quark)]{
    \includegraphics[width=0.97\linewidth]{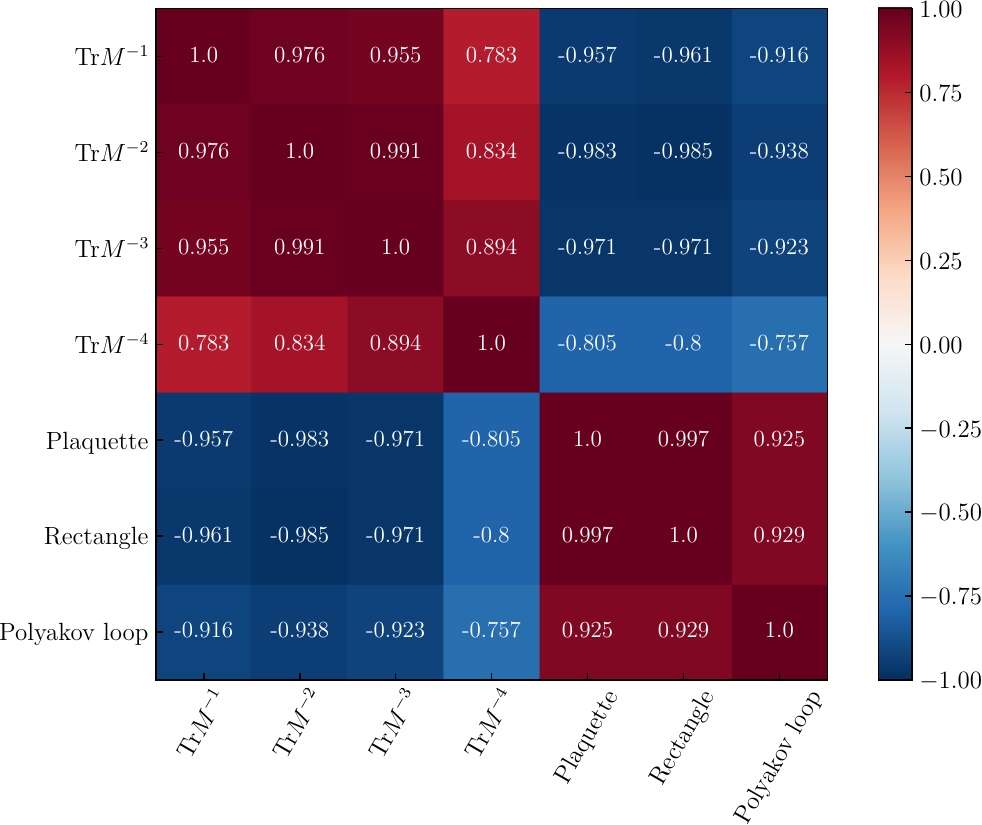}
    \label{fig:subfig-corr-ID-L12T4b1.60k13575}
  }
  \subfigure[$\kappa = 0.13585$, \texttt{L12T4b1.60k13585} (the
    lightest quark)]{
    \includegraphics[width=0.97\linewidth]{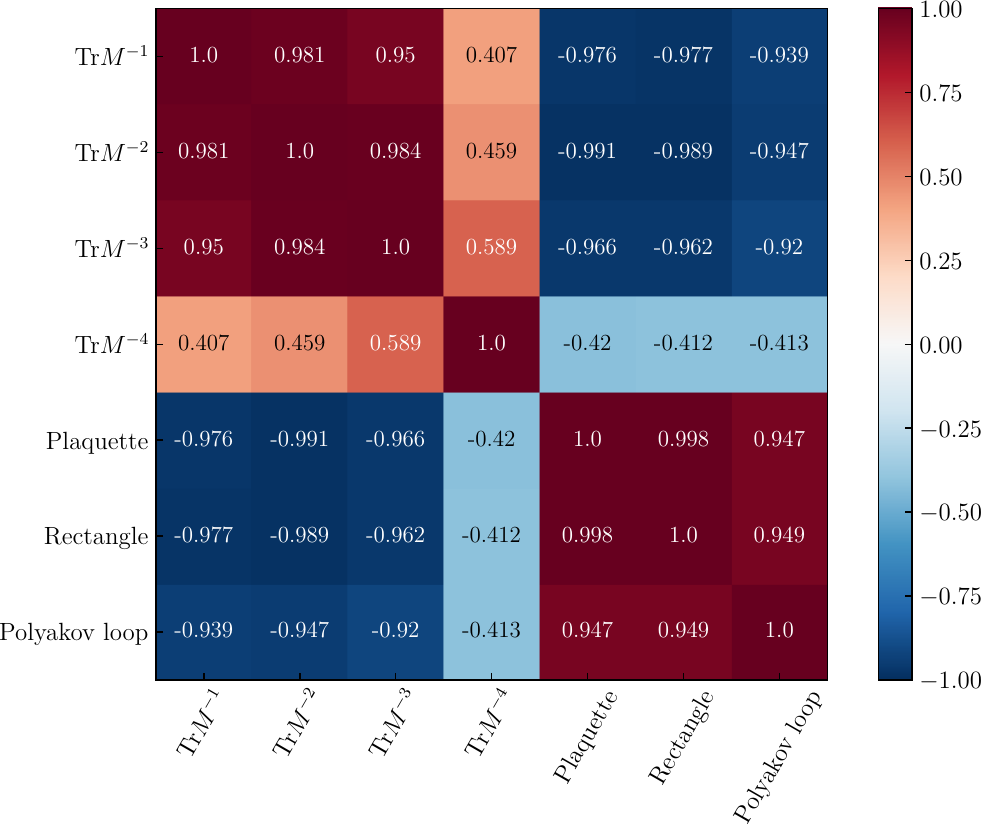}
    \label{fig:subfig-corr-ID-L12T4b1.60k13585}
  }
  \caption{Correlation between physical observables.}
  \label{fig:corr-obs-1}
\end{figure}  
%
For the ML estimation on $Y$ using $X$, the strong correlation between
$X$ and $Y$ is often required.
In Fig.~\ref{fig:corr-obs-1}, we report the correlation between $\Tr
\, M^{-n}$ up to $n=4$, the plaquette, rectangle, and Polyakov loop
from two datasets: \texttt{L12T4b1.60k13575} (the heaviest quark) and
\texttt{L12T4b1.60k13585} (the lightest quark).
The color scale ranges from dark blue for strong negative correlation
($-1$) to dark red for strong positive correlation ($+1$), with white
indicating values around zero where the correlation is weak.
Note that we need to be careful with the $Y = \Tr \, M^{-4}$ case, due
to the weak correlation with other observables.
Except for that, we observe strong correlations in most cases.
Hence, using the overall tendency of strong correlation, we perform
the ML estimation.

\subsection{Two Practical Approaches on ML Estimation}
\label{sec:ML-approach}

In practice, the Dirac operator $M$ is a large and sparse matrix,
making the direct evaluation of the $\Tr \, M^{-n}$ in
Eqs.~(\ref{eq:cond})--(\ref{eq:kurt}) computationally demanding.
To estimate $\Tr \, M^{-n}$, we employ the Hutchinson-type stochastic
trace approximation, in which the trace is evaluated using random
noise vectors as probing sources.
This method substantially reduces the computational cost compared to
an exact trace evaluation, since only a small number of inversions of
the Dirac matrix are required.
Nevertheless, the cost of solving the linear systems by iterative
methods remains considerable, especially when multiple powers of
$M^{-1}$ are involved.

To further reduce this computational expense, we consider the
following two practical approaches:
\begin{description}[
    font=\bfseries, 
    labelsep=0.6em, 
    labelwidth=0pt, 
    widest=ML-1,
    align=left, 
    itemsep=1pt,
  ]
  \ecitem{$\mathcal{F}_{\text{in}}$}{it:trace-feature}
  Here, $\mathcal{F}_{\text{in}}$ denotes the approach in which the ML
  feature set ($\mathcal{F}$) is taken from \emph{internal}
  observables, \textit{i.e.}, trace-based quantities such as $\Tr \,
  M^{-1}$.
  In this approach, $\Tr \, M^{-1}$ obtained from the original data is
  used as the baseline observable, and ML estimations of $\Tr \,
  M^{-n}$ ($n = 2,3,4$) are performed using $\Tr \, M^{-1}$ as an
  input feature.
  The predicted values are then used to construct the cumulants.
  The results obtained from this approach are regarded as the main
  results of the present work.
  \ecitem{$\mathcal{F}_{\text{ex}}$}{it:plaquette-feature}
  Here, $\mathcal{F}_{\text{ex}}$ denotes the approach in which the ML
  feature set ($\mathcal{F}$) is taken from \emph{external}
  observables, \textit{i.e.}, quantities measured during the hybrid
  Monte Carlo (HMC) simulation \cite{Duane:1985hz, Duane:1986iw,
    Duane:1987de} such as the plaquette and rectangle.
  Since the Iwasaki gauge action~\cite{Iwasaki:1985we,
    Iwasaki:1983iya} contains not only the plaquette term but also the
  rectangle term, we include both as features.
  In this approach, correlations between these basic observables and
  the target quantities $\Tr \, M^{-n}$ are utilized.
  All $\Tr \, M^{-n}$ are treated as target observables for ML
  estimation, and the resulting predictions are combined to compute
  the cumulants.
  In this work, the plaquette and rectangle serve as the input
  features for this approach.
\end{description}
Several points should be clarified here.

First, in the \ref{it:trace-feature} approach, the quantity
$\Tr\,M^{-1}$ is taken directly from the original measurements when
constructing the cumulants.
As inferred from Eqs.~\eqref{eq:Q_1}--\eqref{eq:kurt}, the
contribution from $\Tr\,M^{-1}$ is dominant among the trace terms.
Consequently, even for small values of $\mathcal{R}_{\text{LB}}$ [see
  Eq.~\eqref{eq:R-LB-1}], the ML estimates remain similar to those
obtained from the original data (see
Fig.~\ref{fig:kurt-13585-heatmap-ML1-LBP-1-25}, presented later).
This indicates that, although ML estimations for $\Tr\,M^{-n}$
($n=2,3,4$) are present, the cumulant behavior is largely governed by
$\Tr\,M^{-1}$; hence, it is difficult in this setup to disentangle
intrinsic variations of ML accuracy with respect to
$\mathcal{R}_{\text{LB}}$ and $\mathcal{R}_{\text{TR}}$.

Another important aspect of the \ref{it:trace-feature} approach is
that $\Tr\,M^{-1}$ must still be computed by conventional
Dirac-operator inversions during the measurement stage.
Therefore, only the computation of $\Tr\,M^{-n}$ ($n=2,3,4$) benefits
from the proposed reduction.  Because cumulants up to kurtosis require
powers up to fourth order, retaining the complete $\Tr\,M^{-1}$
measurement sets a nominal solve-count floor near $25\%$ of the
baseline, before accounting for training and analysis overhead.

To complement the \ref{it:trace-feature} approach, we also adopt the
\ref{it:plaquette-feature} approach.
In this work, the plaquette and rectangle are used as the input
observables (features) for the \ref{it:plaquette-feature} approach.
The original data used here were first measured for
Ref.~\cite{Ohno:2018gcx}, which employed the Iwasaki gauge action;
accordingly, both the plaquette and rectangle were recorded during the
HMC updates.
It is well known that using multiple correlated features generally
improves the expressive power of supervised models, so we employ these
two features for the \ref{it:plaquette-feature} approach.
Although the Polyakov loop is also available from the HMC stage,
Fig.~\ref{fig:corr-obs-1} shows that its correlation with the
$\Tr\,M^{-n}$ is weaker than that of the plaquette and rectangle.
In fact, reasonable ML estimates can already be achieved when using
only the plaquette and rectangle as input features.
For these reasons, we adopt this two-observable feature set as the
default for the \ref{it:plaquette-feature} approach.

In the \ref{it:plaquette-feature} approach, all $\Tr\,M^{-n}$ are
treated as ML targets so that the precision of the resulting cumulants
fully reflects ML estimation quality.
This setup allows a systematic study of estimation accuracy as a
function of $\mathcal{R}_{\text{LB}}$ and $\mathcal{R}_{\text{TR}}$,
and permits a descriptive study of how agreement changes when all four
trace observables are predicted from smaller labeled subsets.
Although the main results in this work are based on the
\ref{it:trace-feature} approach, a complementary analysis using the
\ref{it:plaquette-feature} approach is included for comparison and
validation.

\subsection{Treatment of Statistical Errors}
\label{sec:stat-err}

The data used in this study for estimating the cumulants of the chiral
condensate are generated in the vicinity of a first-order phase
transition point and are mutually dependent; applying a delete-1
jackknife resampling or \textit{i.i.d.}~bootstrap resampling under an
independence assumption that ignores this dependence may underestimate
the statistical error.

In view of this difficulty, the use of either a delete-$g$ ($g>1$)
jackknife resampling or a block bootstrap resampling could be
considered appropriate methods for estimating the statistical
uncertainties in dependent data.
However, as Eq.~\eqref{eq:P1-1} indicates, the ML estimator in this
work combines statistics from subsets of unequal sizes---the unlabeled
set, bias correction set, and labeled set---so defining a common
resampling unit and constructing synchronized jackknife replicates
across all subsets is nontrivial in practice.
We therefore adopt a block bootstrap resampling to mitigate the
underestimation of statistical errors due to dependence.
The resulting uncertainties are conditional on the empirical block
length selected in Appendix~\ref{sec:block-bs}; they do not include
variability from alternative data partitions or model retraining.

A related discussion of the block bootstrap in the lattice QCD
literature can be found in Ref.~\cite{Christ:2024nxz}.

\subsection{Evaluation Criteria for ML Estimation}

To assess the utility of this approach, we compare the primary ML
estimate $\{\bar{Y}_{\mathcal{P}1}, \sigma_{\mathcal{P}1}\}$ with the
original full-data estimate $\{\bar{Y}_\text{OG}, \sigma_\text{OG}\}$.
We adopt the following evaluation criteria (\textbf{EC-}$\bm{x}$ where
$x=1,2,3$).
\begin{enumerate}[
    label=\textbf{EC-\arabic*},
    ref=\textbf{EC-\arabic*},
    itemsep=0.5pt
  ]
\item \label{it:EC-1} 
  \textbf{Consistency of means.}  This criterion evaluates the degree
  of agreement between the original and ML-estimated sample means.
  we define an overlap metric
  \begin{align}
    x &= \frac{\lvert \bar{Y}_\text{OG}-\bar{Y}_{\mathcal{P}1}
      \rvert}{\sigma_{\text{OG}}} \,,
    \label{eq:x-1}
  \end{align}
  which quantifies the normalized mean separation in units of the
  original uncertainty.
  Values $x \lesssim 1$ indicate sufficient overlap within one
  standard deviation, whereas $x \gtrsim 1$ signals that the
  ML-estimated sample mean deviates beyond the $1\sigma$ range of the
  original sample mean.

\item \label{it:EC-2}
  \textbf{Error ratio near unity.}  We assess whether
  \begin{align}
    r &= \frac{\sigma_{\mathcal{P}1}}{\sigma_{\text{OG}}}
    \label{eq:r-1}
  \end{align}
  is close to unity, \textit{i.e.}, $r \approx 1$, meaning that the ML
  estimation uncertainty $\sigma_{\mathcal{P}1}$ is comparable to the
  original full-data uncertainty $\sigma_\text{OG}$.

\item \label{it:EC-3}
  \textbf{Gaussian Bhattacharyya overlap.}  As a complementary
  descriptive summary that jointly reflects mean separation and
  relative uncertainty under a Gaussian approximation, we evaluate
  the Bhattacharyya coefficient
  $C_{\text{B}}(x,r)$ defined in Eq.~\eqref{eq:bha-coeff-1} with $x$
  and $r$ given in Eqs.~\eqref{eq:x-1} and \eqref{eq:r-1}.
  Values near $C_{\text{B}}(x,r) \simeq 0.95$ are used as a visual
  guide to close agreement, rather than as a universal pass--fail
  threshold; examples are provided below (see
  Table~\ref{tab:cb-thresholds-1}).
\end{enumerate}
If an ML estimator receives $x \lesssim 1$ in \ref{it:EC-1} and
satisfies $r \approx 1$ in \ref{it:EC-2}, we interpret this as being
consistent with a close reproduction of the corresponding original
result.
We also report the metric in \ref{it:EC-3} as a complementary summary
of overlap.
Note that \ref{it:EC-1} and \ref{it:EC-2} are assessed independently.
While both are informative, they probe mean consistency and relative
uncertainty, respectively, so a method may score well on one but
poorly on the other, leaving the overall assessment ambiguous.
For example, two estimates may agree in mean within $1\sigma$ while
one has a much larger uncertainty, or they may have comparable
uncertainties but exhibit an appreciable mean shift that reduces
overlap.
Hence, to capture both effects simultaneously with a single,
interpretable metric, we additionally employ the Bhattacharyya
coefficient $C_{\text{B}}(x,r)$ (\ref{it:EC-3}), defined by
\begin{align}
  C_{\text{B}}(x,r) &= \sqrt{\frac{2r}{1+r^{2}}}\,
  \exp\!\left[-\frac{x^{2}}{4(1+r^{2})}\right] \,,
  \label{eq:bha-coeff-1}
\end{align}
where $x$ and $r$ are defined in Eqs.\eqref{eq:x-1} and
\eqref{eq:r-1}.
The Gaussian Bhattacharyya coefficient $C_{\text{B}}(x,r)$ takes
values in $[0,1]$: values close to $1$ indicate strong overlap between
the two Gaussian summaries, whereas values close to $0$ indicate
little overlap.
A key feature of $C_{\text{B}}(x,r)$ is that it depends on both the
means and the uncertainties of the two distributions: it decreases not
only as the separation between the means grows, but also when the
means are close, yet one uncertainty is significantly larger than the
other.

To build intuition for the Bhattacharyya coefficient, consider a few
examples.
When $x=1$ and $r=1$, the original and ML estimates have identical
uncertainties, and their means are separated by exactly one standard
deviation of the original.
In this case $C_{\text{B}}(1,1)=e^{-1/8}\approx 0.882$.

Now suppose the ML estimation uncertainty is about $1.1$ times the
original ($r=1.1$) while keeping $x=1$.
The coefficient increases slightly to $C_{\text{B}}(1,1.1)\approx
0.891$, reflecting the greater overlap due to the broader ML
estimation uncertainty.
By contrast, if $r=0.9$ with $x=1$, $C_{\text{B}}(1,0.9)\approx
0.869$, indicating reduced overlap.

\begin{table}[b!]
  \renewcommand{\arraystretch}{1.2}
  \caption{Admissible ranges of $r$ (at $x=0$) and $x$ (at $r=1$) for
    selected Bhattacharyya-coefficient thresholds $C_{\text{B}}\ge c$.
    The coefficient $C_{\text{B}}(x,r)$ is defined in
    Eq.~\eqref{eq:bha-coeff-1}, with $x$ and $r$ defined in
    Eqs.~\eqref{eq:x-1} and \eqref{eq:r-1}.
    At $x=0$, $C_{\text{B}}(0,r)=C_{\text{B}}(0,1/r)$, so feasible $r$
    intervals appear as reciprocal pairs.}
  \label{tab:cb-thresholds-1}
  \begin{ruledtabular}
    \begin{tabular}{@{\quad}c@{\quad}|@{\quad}c@{\quad}|@{\quad}c@{\quad}}
      $c_\text{th}$ & Feasible $r$ ($x=0$) & Feasible $x$ ($r=1$) \\
      \hline
      $0.95$ & $[\,0.631,\;1.585\,]$ & $[\,0,\;0.641\,]$ \\
      $0.96$ & $[\,0.664,\;1.506\,]$ & $[\,0,\;0.571\,]$ \\
      $0.97$ & $[\,0.703,\;1.423\,]$ & $[\,0,\;0.494\,]$ \\
      $0.98$ & $[\,0.751,\;1.331\,]$ & $[\,0,\;0.402\,]$ \\
      $0.99$ & $[\,0.818,\;1.223\,]$ & $[\,0,\;0.284\,]$ \\
    \end{tabular}
  \end{ruledtabular}
\end{table}
%
As a reference, Table~\ref{tab:cb-thresholds-1} lists, for thresholds
$c_\mathrm{th}\in\{0.95,0.96,0.97,0.98,0.99\}$, the feasible ranges of
$r$ when $x=0$ and of $x$ when $r=1$, \textit{i.e.}, the idealized
cases in which either the mean separation vanishes or the relative
uncertainty equals unity.
In practice, however, both $x$ and $r$ generally deviate from these
ideals, and since $C_{\text{B}}(x,r)$ depends jointly on mean
separation and relative uncertainty, the table entries should be read
as guidance rather than used mechanistically as binary pass/fail
rules.
While precise tolerances are context-dependent, coefficients in the
vicinity of $C_{\text{B}}(x,r) \approx 0.95$ or higher indicate
substantial overlap in this Gaussian summary.  We therefore use this
value only as an exploratory visual guide to ``close agreement'' in
the present fixed-dataset study.

Accordingly, when discussing the similarity between the ML estimation
and original full-data results, we use \ref{it:EC-3} (the Gaussian
Bhattacharyya overlap) as a compact descriptive metric.
When interpreting the resulting $C_{\text{B}}(x,r)$
values---\textit{e.g.}, to disentangle the roles of mean separation
and relative uncertainty---we consult \ref{it:EC-1} and \ref{it:EC-2}
as ancillary diagnostics.
This choice is motivated by the fact that \ref{it:EC-1} and
\ref{it:EC-2} probe mean consistency and relative uncertainty
separately, which can lead to ambiguous assessments, whereas the
Bhattacharyya coefficient provides a single interpretable measure that
captures both effects simultaneously.
It offers a compact way to display mean separation and relative
uncertainty in a single scalar quantity.  It should not, however, be
interpreted as a test of statistical equivalence or as accounting for
the covariance between the ML and full-data estimates, which are
constructed from overlapping information.

\section{Trace estimation}
\label{sec:trace}

\begin{figure}[b!]
  \includegraphics[width=0.97\linewidth]{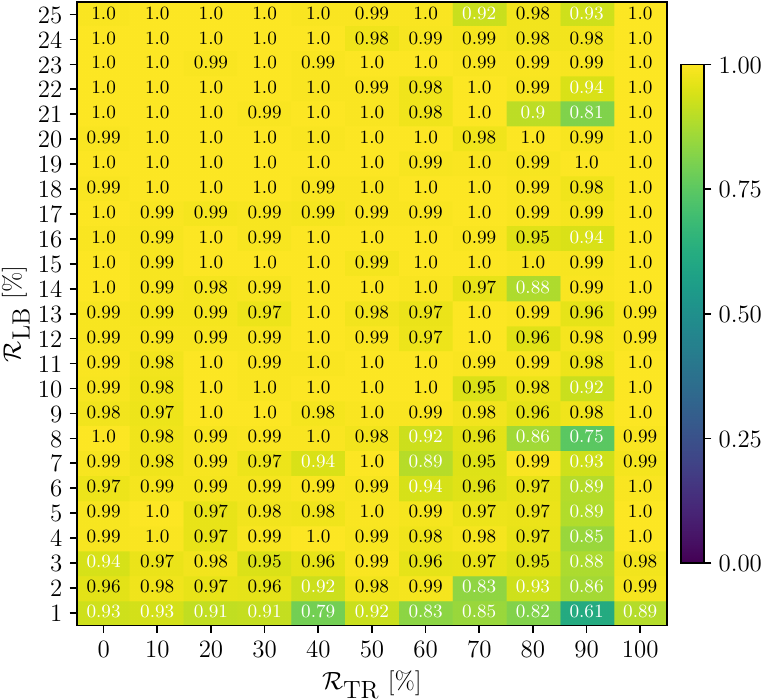}
  \caption{Bhattacharyya coefficient $C_\text{B}$ [\ref{it:EC-3}] map
    for the primary estimator $\mathcal{P}1$ in the $\Tr\,M^{-4}$
    estimation performed using $\Tr\,M^{-1}$ as the input feature in
    the \ref{it:trace-feature} approach, corresponding to the dataset
    with ID \texttt{L12T4b1.60k13585}.}
  \label{fig:trace-heatmap-ML1-LBP-1-25}
\end{figure}

\begin{figure}[b!]
  \includegraphics[width=0.97\linewidth]{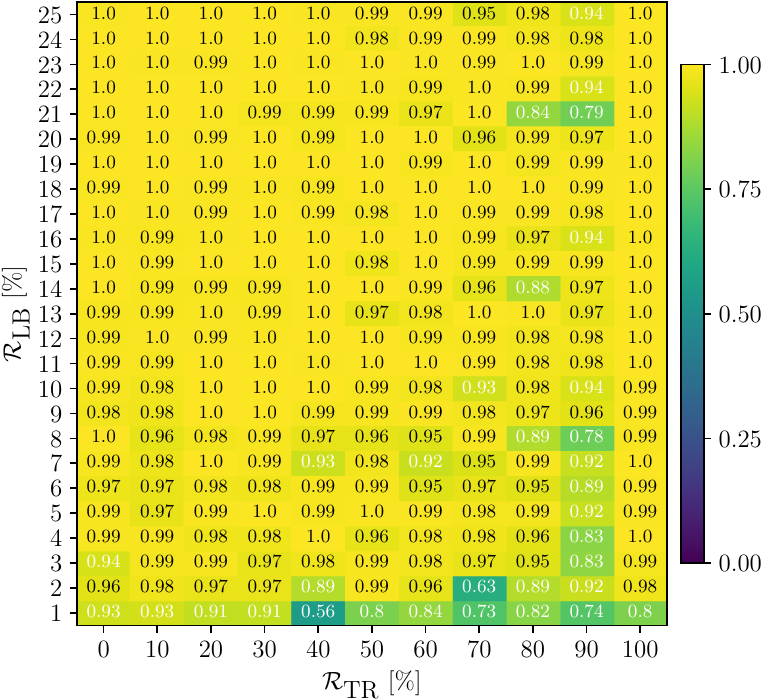}
  \caption{Bhattacharyya coefficient $C_\text{B}$ map for the primary
    estimator $\mathcal{P}1$ in the $\Tr\,M^{-4}$ estimation performed
    using the plaquette and rectangle as input features in the
    \ref{it:plaquette-feature} approach, corresponding to the dataset
    with ID \texttt{L12T4b1.60k13585}.
    The same plotting conventions as in
    Fig.~\ref{fig:trace-heatmap-ML1-LBP-1-25} are used.}
  \label{fig:trace-heatmap-ML2-LBP-1-25}
\end{figure}

In this section, we describe the ML estimation of the $\Tr \, M^{-n}$.
Although part of this trend can already be seen in
Fig.~\ref{fig:corr-obs-1}, the detailed correlation analysis reveals
that correlations among observables tend to weaken as $\kappa$
increases (quarks with lighter mass).
Moreover, among the target observables $\Tr\,M^{-n}$, the correlation
with the input features was found to be weakest for $\Tr\,M^{-4}$.
Accordingly, the ensemble with ID \texttt{L12T4b1.60k13585} in
Table~\ref{tab:ensemble-1}, with $\Tr\,M^{-4}$ as the target
observable, provides the most conservative setup for testing the
robustness of the ML estimation.

Fig.~\ref{fig:trace-heatmap-ML1-LBP-1-25} presents the corresponding
results obtained using the \ref{it:trace-feature} approach, in which
$\Tr\,M^{-4}$ is estimated from $\Tr\,M^{-1}$ serving as the input
feature.
The figure shows the Gaussian Bhattacharyya coefficient $C_\text{B}$
map for the primary estimator $\mathcal{P}1$, providing a compact
descriptive measure of the agreement between the ML estimate and the
original full-data reference.
A higher $C_\text{B}$ value indicates stronger overlap between the two
Gaussian summaries under the adopted criteria.

Fig.~\ref{fig:trace-heatmap-ML2-LBP-1-25} shows the corresponding
results obtained with the \ref{it:plaquette-feature} approach, which
employs the plaquette and rectangle as input features.
The same plotting conventions as in
Fig.~\ref{fig:trace-heatmap-ML1-LBP-1-25} are used, allowing a direct
comparison between the two approaches.

As observed in
Fig.~\ref{fig:corr-obs-1}\subref{fig:subfig-corr-ID-L12T4b1.60k13585},
this dataset exhibits relatively weak correlations between
$\Tr\,M^{-4}$ and the other observables compared with those among the
lower-order traces.
In particular, the correlation coefficients between $\Tr\,M^{-4}$ and
$\Tr\,M^{-1}$, plaquette, and rectangle are approximately $0.407$,
$0.420$, and $0.412$, respectively.
Hence, the input features used in the \ref{it:trace-feature} approach
and \ref{it:plaquette-feature} approaches possess nearly identical
levels of correlation with the target observable, which explains why
both approaches yield a comparable level of predictive performance for
this dataset.

The scan results indicate that, for both approaches, the $C_\text{B}$
scores tend to be low in the region where $\mathcal{R}_{\text{LB}} =
1\,\%$ across all scanned values of $\mathcal{R}_{\text{TR}}$.
This tendency gradually diminishes as $\mathcal{R}_{\text{LB}}$
increases: higher $\mathcal{R}_{\text{LB}}$ values lead to a broader
region of $\mathcal{R}_{\text{TR}}$ exhibiting consistently large
$C_\text{B}$ scores, typically starting from smaller
$\mathcal{R}_{\text{TR}}$ values.

Along the same $\mathcal{R}_{\text{LB}}$ contour, there exist a few
isolated points where $C_\text{B}$ accidentally attains relatively
high scores; however, the overall trend shows that higher
$\mathcal{R}_{\text{TR}}$ (equivalently, smaller
$\mathcal{R}_{\text{BC}} = 1 - \mathcal{R}_{\text{TR}}$) regions tend
to yield lower $C_\text{B}$ scores.
When $\mathcal{R}_{\text{LB}}$ is small, consistently high
$C_\text{B}$ values appear only in a narrow range of small
$\mathcal{R}_{\text{TR}}$, whereas as $\mathcal{R}_{\text{LB}}$
increases, this high-$C_\text{B}$ region gradually expands from the
low-$\mathcal{R}_{\text{TR}}$ side, forming a smooth curved pattern in
the map.
Overall, the lowest $C_\text{B}$ values are observed near the region
with the smallest $\mathcal{R}_{\text{LB}}$ and the largest
$\mathcal{R}_{\text{TR}}$, representing the least favorable
configuration in terms of ML estimation consistency.
This overall trend indicates that, when adopting a bias-corrected ML
approach, it is important to secure a sufficiently large bias
correction set from the labeled set in order to maintain better
prediction quality.

\begin{figure}[tb]
  \subfigure[bias correction set]{
    \includegraphics[width=0.97\linewidth]{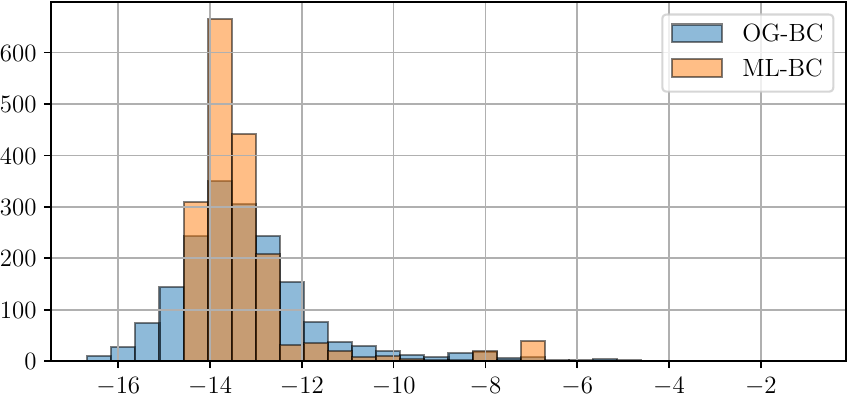}
      \label{fig:subfig-trace-histogram-LBP-1-25-BC}
  }
  \subfigure[unlabeled set]{
    \includegraphics[width=0.97\linewidth]{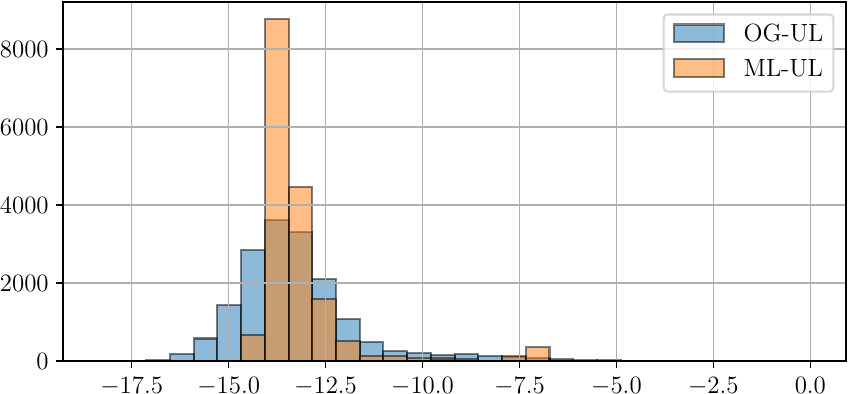}
      \label{fig:subfig-trace-histogram-LBP-1-25-UL}
  }
  \caption{Comparison histograms of the original data (blue) and ML
    predictions (orange) for $\Tr\,M^{-4}$ at
    $\mathcal{R}_{\mathrm{LB}}=15\,\%$ and
    $\mathcal{R}_{\mathrm{TR}}=40\,\%$.
    Panel~\subref{fig:subfig-trace-histogram-LBP-1-25-BC} shows the
    bias-correction set, while
    Panel~\subref{fig:subfig-trace-histogram-LBP-1-25-UL} shows the
    unlabeled set.
    According to Fig.~\ref{fig:trace-heatmap-ML2-LBP-1-25}, this
    parameter choice corresponds to a Bhattacharyya coefficient
    $C_{\mathrm{B}}\simeq 1.0$, indicating strong overlap between the
    corresponding Gaussian summaries.
    The difference between the original data and the ML prediction in
    the bias-correction subset
    [Panel~\subref{fig:subfig-trace-histogram-LBP-1-25-BC}]
    contributes as a bias-correction term which, together with the
    unlabeled-set ML prediction (orange in
    Panel~\subref{fig:subfig-trace-histogram-LBP-1-25-UL}), enters the
    construction of $\bar{Y}_{\mathcal{P}1}$ via Eq.~\eqref{eq:P1-1}.
    Note that in
    Panel~\subref{fig:subfig-trace-histogram-LBP-1-25-UL}, a small
    number of unlabeled samples (21 out of 17\,000) exceed the upper
    plotting range ($x>0.0$) and are omitted for visual clarity,
    whereas all samples in
    Panel~\subref{fig:subfig-trace-histogram-LBP-1-25-BC} lie within
    the displayed range. }
  \label{fig:trace-histogram-LBP-1-25}
\end{figure}

To illustrate what a high Bhattacharyya overlap corresponds to at the
distribution level, we also present a direct comparison at a
representative point in the scan.
Fig.~\ref{fig:trace-histogram-LBP-1-25} compares the original and
ML-predicted distributions for two subsets at
$\mathcal{R}_{\mathrm{LB}}=15\,\%$ and
$\mathcal{R}_{\mathrm{TR}}=40\,\%$:
panel~\subref{fig:subfig-trace-histogram-LBP-1-25-BC} shows the
bias-correction set, while
panel~\subref{fig:subfig-trace-histogram-LBP-1-25-UL} shows the
unlabeled set.
In each panel, the original data (OG) and the ML prediction (ML) are
overlaid as histograms.
These subsets enter the construction of the $\mathcal{P}1$ estimator
$\bar{Y}_{\mathcal{P}1}$ according to Eq.~\eqref{eq:P1-1}, and the
agreement of the resulting prediction with the original distribution
is quantified by the Bhattacharyya coefficient $C_{\mathrm{B}}$ shown
in the heatmap.
At this $(\mathcal{R}_{\mathrm{LB}},\mathcal{R}_{\mathrm{TR}}) =
(15\,\%,40\,\%)$ point, the heatmap gives $C_{\mathrm{B}}\simeq 1$,
corresponding to strong overlap between the Gaussian summaries of the
predicted and original estimates.
The subset-level histograms clarify this behavior: although the
original and ML distributions are not strictly identical within each
subset, they exhibit broadly consistent central values and comparable
spread.
Consequently, once combined through Eq.~\eqref{eq:P1-1}, the resulting
$\mathcal{P}1$-level distribution achieves a near-unity Bhattacharyya
overlap, consistent with the highest-$C_{\mathrm{B}}$ region of the
map.

It is also worth noting that, in the columns corresponding to
$\mathcal{R}_{\text{TR}} = 100\,\%$ in
Figs.~\ref{fig:trace-heatmap-ML1-LBP-1-25} and
\ref{fig:trace-heatmap-ML2-LBP-1-25} ---where no bias correction set
is used and the entire labeled set is devoted to training---the
$C_\text{B}$ scores remain almost uniformly high, unlike those at
$\mathcal{R}_{\text{TR}} = 90\,\%$ just beside.
For the fixed dataset studied here, the trace-estimation results alone
therefore do not provide a uniformly clear advantage for reserving a
bias-correction set over the pure-training setup in terms of
$C_\text{B}$.

However, as will be shown later in the context of the ML estimation of
cumulants, the situation changes markedly: when no bias correction is
applied, the Bhattacharyya overlap between the ML estimation and
original full-data results can decrease substantially, leading to a
qualitatively different behavior within the fixed dataset.

\section{Cumulant estimation at the single ensemble}
\label{sec:cumulant-single}

\begin{figure}[tb]
  \includegraphics[width=0.97\linewidth]{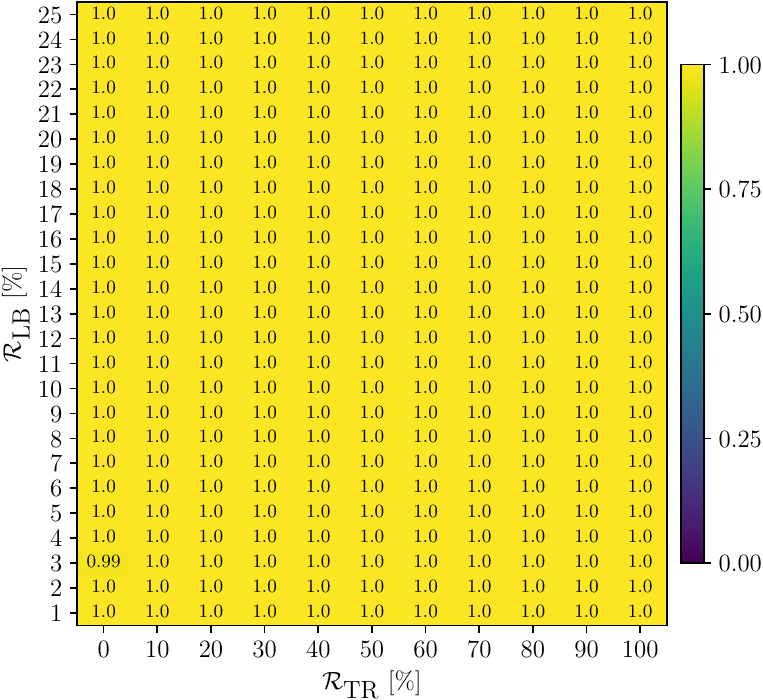}
  \caption{Bhattacharyya coefficient $C_\text{B}$ map for the primary
    estimator $\mathcal{P}1$ in the kurtosis estimation, performed
    using $\Tr\,M^{-1}$ from the original data and $\Tr\,M^{-n}$
    ($n=2,3,4$) predicted from $\Tr\,M^{-1}$ as the input feature in
    the \ref{it:trace-feature} approach, based on the dataset with ID
    \texttt{L12T4b1.60k13585}.}
  \label{fig:kurt-13585-heatmap-ML1-LBP-1-25}
\end{figure}

\begin{figure}[tb]
  \includegraphics[width=0.97\linewidth]{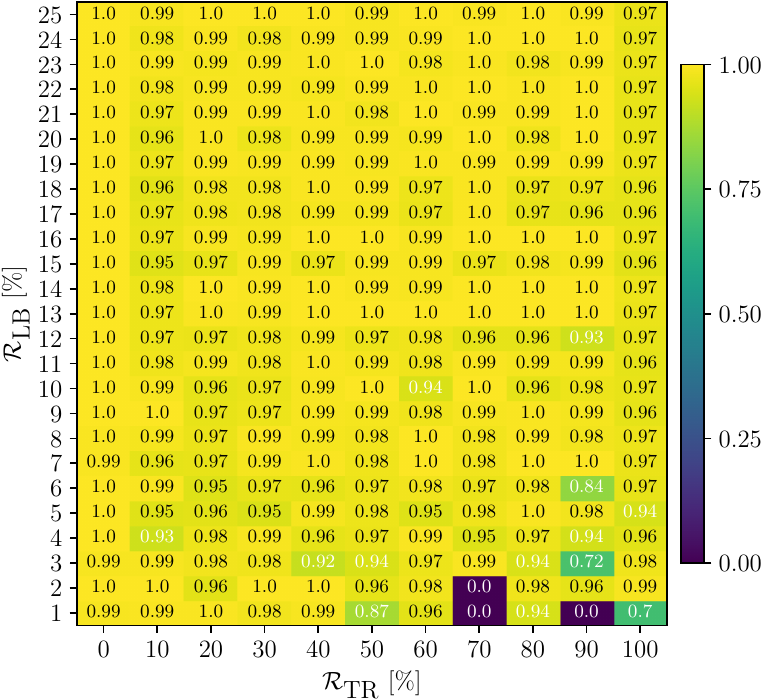}
  \caption{Bhattacharyya coefficient $C_\text{B}$ map for the primary
    estimator $\mathcal{P}1$ in the kurtosis estimation, performed
    using plaquette and rectangle as input features in the
    \ref{it:plaquette-feature} approach, based on the dataset with ID
    \texttt{L12T4b1.60k13585}.
    The same plotting conventions as in
    Fig.~\ref{fig:kurt-13585-heatmap-ML1-LBP-1-25} are used.}
  \label{fig:kurt-13585-heatmap-ML2-LBP-1-25}
\end{figure}

In this section, we discuss the procedure for estimating the cumulants
by combining the trace observables $\Tr\,M^{-n}$ obtained using the ML
methods described in Sec.~\ref{sec:trace}.
While the cumulants can, in principle, be computed directly from
Eqs.~\eqref{eq:Q_1}--\eqref{eq:kurt}, additional considerations are
required when the trace inputs are provided by bias-corrected ML
estimations, as explained below.

In this work, the bias-corrected ML estimation scheme introduced in
Sec.~\ref{sec:framework} [see Eq.~\eqref{eq:P1-1}] is adopted as the
main framework.
The detailed construction of the subsets $S^Z_{\text{LB}}$,
$S^Z_{\text{UL}}$, $S^Z_{\text{TR}}$, and $S^Z_{\text{BC}}$ has
already been described in Sec.~\ref{sec:dataset-split}; here we only
summarize the aspects relevant to the cumulant estimation.
When computing the cumulants, the same subsets as defined in the trace
estimation stage are employed—namely $S^{P}_{\text{UL}}$,
$S^Y_{\text{BC}}$, $S^{P}_{\text{BC}}$, and $S^Y_{\text{LB}}$—so that
all relevant correlations are consistently preserved.

In the conventional (non-ML) approach, the moments $Q_1$–$Q_4$ defined
in Eqs.~\eqref{eq:Q_1}--\eqref{eq:Q_4} are first computed
configuration by configuration from $\Tr\,M^{-n}$ without resampling,
and resampling is performed only afterward when forming the ensemble
averages.
Since that is the stage where resampling becomes necessary, we follow
the same principle in the ML estimation as well.

In the bias-corrected ML estimation, the evaluation of
Eq.~\eqref{eq:P1-1} requires replicas of each subset, because
$Q_1$–$Q_4$ are constructed from combinations of $Y^{P}_{\text{UL}}$,
$Y_{\text{BC}}$, $Y^{P}_{\text{BC}}$, and $Y_{\text{LB}}$.
Therefore, bootstrap resampling is performed separately for each
subset to generate the necessary replicas, after which $Q_1$–$Q_4$ are
computed for each bootstrap sample according to Eq.~\eqref{eq:P1-1}
for the primary estimator $\mathcal{P}1$.
The ensemble of these replicas is then used to obtain the mean and
statistical uncertainty of the cumulants, ensuring consistency with
the resampling procedure adopted in the non-ML case.

Fig.~\ref{fig:kurt-13585-heatmap-ML1-LBP-1-25} presents the ML
estimation results for the kurtosis obtained from the
\ref{it:trace-feature} approach using the same dataset
(\texttt{L12T4b1.60k13585}) as in the previous section
    [Sec.~\ref{sec:trace}].
Across the entire scanned range of $\mathcal{R}_{\text{LB}}$ and
$\mathcal{R}_{\text{TR}}$, the $C_\text{B}$ values remain close to
unity ($C_\text{B} \approx 1.0$), indicating strong overlap between
the Gaussian summaries of the ML and original estimates.
One plausible interpretation is that the cumulants are dominated by
the contribution of $\Tr\,M^{-1}$ in
Eqs.~\eqref{eq:Q_1}--\eqref{eq:kurt}.
Although higher-order traces $\Tr\,M^{-n}$ ($n=2,3,4$) contribute
subdominantly to the magnitude, they can still affect the cumulants
through the nonlinear moment and cumulant combinations; accordingly,
small residual biases in these components may become more visible at
the cumulant stage.
Since the \ref{it:trace-feature} approach directly uses the original
$\Tr\,M^{-1}$ as both $X$ and $Y$, the intrinsic fluctuations of the
original $\Tr\,M^{-1}$ data propagate almost unchanged into the
cumulant calculation.
Within the fixed dataset studied here, this shared dominant component
accounts for much of the observed overlap in the
\ref{it:trace-feature} approach.

\begin{figure*}[tb]
  \centering
  \includegraphics[width=0.97\linewidth]{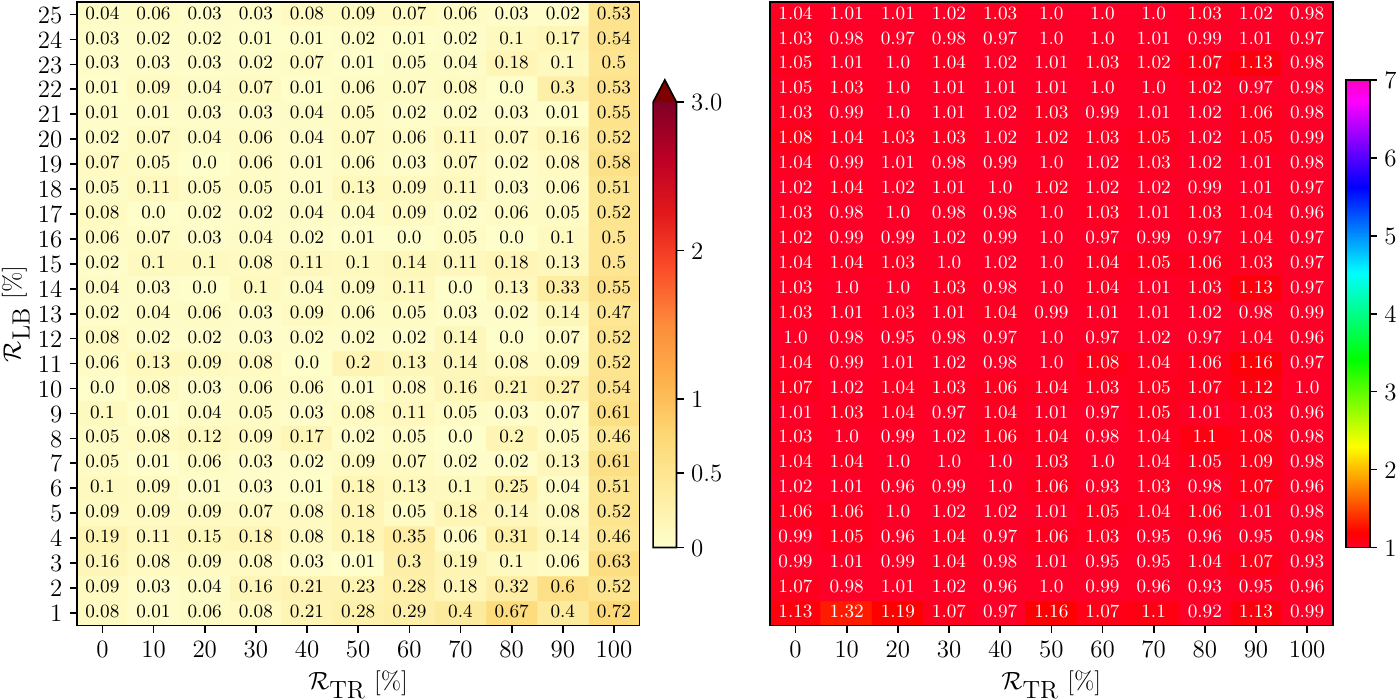}
  \caption{Heatmap showing the results of \ref{it:EC-1} and
    \ref{it:EC-2} evaluations for the kurtosis estimation based on the
    \ref{it:plaquette-feature} approach, corresponding to the dataset
    with ID \texttt{L12T4b1.60k13580} ($\mathcal{P}1$).
    Within the panel, the left half represents the \ref{it:EC-1}
    results, while the right half corresponds to \ref{it:EC-2}.
    In the \ref{it:EC-1} map, lighter shades indicate smaller
    normalized mean separations $x$ (defined in Eq.~\eqref{eq:x-1}),
    corresponding to better agreement between the original and
    predicted means, whereas darker shades represent larger $x$
    values, signaling poorer overlap.
    The kurtosis is computed using $\Tr\,M^{-n}$ predicted from
    plaquette and rectangle as the input feature.}
  \label{fig:kurt-13580-heatmap-ML2-LBP-1-25-P1-EC1-EC2}
\end{figure*}

\begin{figure}[b!]
  \includegraphics[width=0.97\linewidth]{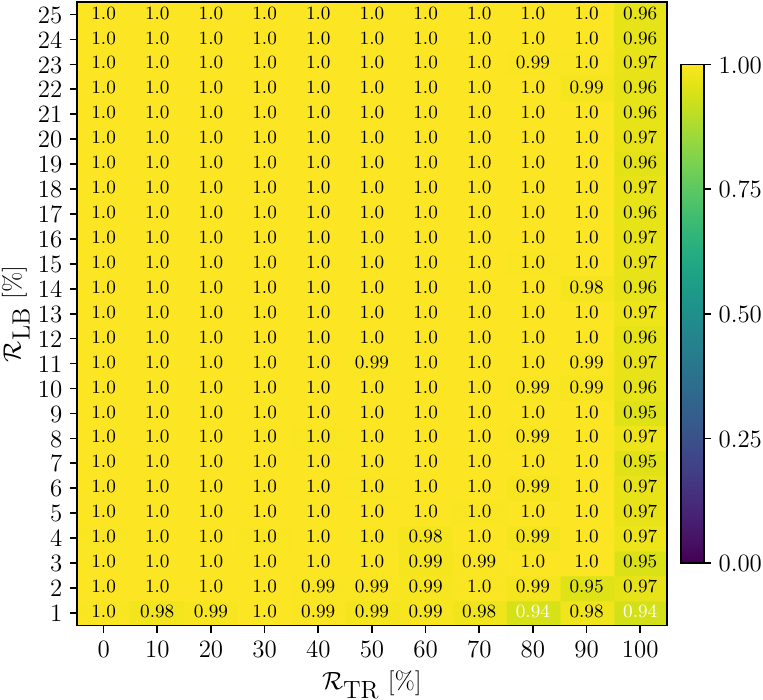}
  \caption{Bhattacharyya coefficient $C_\text{B}$ map for the primary
    estimator $\mathcal{P}1$ in the kurtosis estimation, performed
    using $\Tr\,M^{-n}$ predicted from the plaquette and rectangle
    as input features in the \ref{it:plaquette-feature} approach,
    corresponding to the dataset with ID
    \texttt{L12T4b1.60k13580}.}
  \label{fig:kurt-13580-heatmap-ML2-LBP-1-25}
\end{figure}

To examine the extent to which cumulants can be estimated by ML
estimations alone, we also apply the \ref{it:plaquette-feature}
approach, whose results are shown in
Fig.~\ref{fig:kurt-13585-heatmap-ML2-LBP-1-25}.
Here, as in the case of $\Tr\,M^{-4}$ estimation discussed earlier,
one again observes that regions with small $\mathcal{R}_{\text{LB}}$
and large $\mathcal{R}_{\text{TR}}$ (equivalently, small
$\mathcal{R}_{\text{BC}} = 1 - \mathcal{R}_{\text{TR}}$) tend to yield
lower $C_\text{B}$ scores.
Using $C_\text{B}\approx0.95$ as an exploratory visual guide,
high-$C_\text{B}$ regions begin to emerge at approximately
$\mathcal{R}_{\text{LB}} \gtrsim 5\,\%$ for this fixed dataset.

Another noteworthy feature is seen in the column corresponding to
$\mathcal{R}_{\text{TR}} = 100\,\%$, where no bias correction is
applied and the entire labeled set is used for training.
In contrast to the neighboring $\mathcal{R}_{\text{TR}} = 90\,\%$
column, the $C_\text{B}$ values there are slightly lower, indicating a
mild degradation in overlap quality.

To further illustrate this tendency,
Fig.~\ref{fig:kurt-13580-heatmap-ML2-LBP-1-25} shows the kurtosis
estimation for the ensemble \texttt{L12T4b1.60k13580} under the
\ref{it:plaquette-feature} approach.
This ensemble, among the five in Table \ref{tab:ensemble-1}, is the
one closest to the first-order transition, exhibiting strong
autocorrelation effects.
In the present scan, it achieves nearly uniform high $C_\text{B}$
values across the entire $(\mathcal{R}_{\text{LB}},
\mathcal{R}_{\text{TR}})$ scan, even under the
\ref{it:plaquette-feature} approach.
However, even in this case, while all other columns yield strong
Gaussian-summary overlap ($C_\text{B} \approx 1.0$), the
$\mathcal{R}_{\text{TR}} = 100\,\%$ column shows slightly lower scores
($C_\text{B} \approx 0.96$), close to the exploratory guide value of
$C_\text{B} \approx 0.95$.

To further investigate why the $C_\text{B}$ scores in the
$\mathcal{R}_{\text{TR}} = 100\,\%$ column are slightly lower than
those in the bias-corrected regions, we examine
Fig.~\ref{fig:kurt-13580-heatmap-ML2-LBP-1-25-P1-EC1-EC2}, which
displays the \ref{it:EC-1} and \ref{it:EC-2} diagnostics for the
primary estimator $\mathcal{P}1$.
The left panel represents the \ref{it:EC-1} results.
In most bias-corrected regions, the normalized mean separation remains
as small as $x \lesssim 0.1$, implying $\lvert \bar{Y}_\text{OG} -
\bar{Y}_{\mathcal{P}1} \rvert \lesssim 0.1 \, \sigma_\text{OG}$.
By contrast, in the $\mathcal{R}_{\text{TR}} = 100\,\%$ column, the
deviation typically increases to $x \gtrsim 0.5$, corresponding to
$\lvert \bar{Y}_\text{OG} - \bar{Y}_{\mathcal{P}1} \rvert \gtrsim 0.5
\, \sigma_\text{OG}$.
Although a $\sim 0.5 \, \sigma$ discrepancy could still be regarded as
acceptable in general terms, the key observation is that the
bias-corrected estimators achieve nearly uniform agreement within $0.1
\, \sigma$, highlighting the comparatively inferior overlap in the
$\mathcal{R}_{\text{TR}} = 100\,\%$ case.
This deviation naturally explains the slightly reduced $C_\text{B}$
values observed in that column.

Figs.~\ref{fig:kurt-13585-heatmap-ML2-LBP-1-25}
and~\ref{fig:kurt-13580-heatmap-ML2-LBP-1-25} show that, in the
\ref{it:plaquette-feature} approach, the contrast between corrected
and uncorrected estimates becomes more pronounced at the cumulant
stage.
While this effect was less visible in the direct estimation of
$\Tr\,M^{-n}$, small residual biases in the uncorrected $\Tr\,M^{-n}$
accumulate through the moment and cumulant computations, leading to
the noticeable $C_\text{B}$ reduction observed at the end.
With the aid of
Fig.~\ref{fig:kurt-13580-heatmap-ML2-LBP-1-25-P1-EC1-EC2}, this trend
can be further understood as the consequence of the increased mean
separation seen in the $\mathcal{R}_{\text{TR}}=100\,\%$ case.
Within this fixed dataset, these comparisons indicate that residual
biases, though seemingly negligible at the $\Tr\,M^{-n}$ level, can
propagate and become amplified through higher-order moments and
cumulant combinations.

\section{Cumulant estimation with multi-ensemble reweighting}
\label{sec:cumulant-reweighting}

In the previous section, we discussed the ML estimation of cumulants
for a single ensemble characterized by a specific lattice volume $V$,
gauge coupling $\beta$, and hopping parameter $\kappa$.
In this section, we extend the discussion to multi-ensemble
reweighting, in which multiple ensembles share the same $V$ and
$\beta$ but have different $\kappa$ values that are analyzed
collectively, incorporating the ML estimation into the procedure.

The multi-ensemble reweighting was first introduced in
Ref.~\cite{Ferrenberg:1988yz, Ferrenberg:1989ui}.
Since then, it has been widely employed in the lattice QCD community.
For instance, Ref.~\cite{Jin:2015taa} investigated the phase structure
of $N_{\text{f}} = 3$ QCD, while Ref.~\cite{Kuramashi:2016kpb}
explored that of $N_{\text{f}} = 2+1$ QCD.
The dataset used in the present study was originally generated for
Ref.~\cite{Ohno:2018gcx}, which investigated the phase structure of
$N_{\text{f}} = 4$ QCD.

The complete multi-ensemble reweighting procedure adopted in this work
is summarized in Appendix~\ref{sec:app-multi-rw}.
For readability, we only provide a high-level roadmap here and
introduce four \emph{index-aligned} datasets
$\mathcal{S}_1$--$\mathcal{S}_4$ for later reference.
Their explicit definitions are given in
Appendix~\ref{sec:app-multi-rw}
[Eqs.~\eqref{eq:EA-1}--\eqref{eq:EA-4}], but conceptually
$\mathcal{S}_1$ corresponds to the fully original dataset,
$\mathcal{S}_2$ replaces only the unlabeled sector with ML
predictions, and $\mathcal{S}_3$--$\mathcal{S}_4$ are labeled sector
variants used to preserve the intended role of bias correction without
structural contamination.

At a conceptual level, the procedure proceeds through four main stages
(see also Appendix~\ref{sec:app-multi-rw}).
First, the datasets are reconstructed with strict index alignment so
that the gauge configuration ordering matches that of the original
measurements.
Next, the free-energy offsets are determined using the Newton--Raphson
method, enabling the evaluation of the reweighting factors
$w_{a;n}(\kappa_r)$ along the $\kappa$ trajectory.
The traces $\Tr\,M^{-j}$ are then approximately shifted to the target
$\kappa_r$ and converted into $Q_j$ and $w\,Q_j$.
Finally, bootstrap replicas are generated, and the transition point
$\kappa_t$ is obtained through interpolation.

While the detailed step-by-step description is deferred to
Appendix~\ref{sec:app-multi-rw}, we emphasize several structural
points below that are essential for understanding the physical
results.

A key aspect of the procedure is that each ensemble is first
reconstructed by strictly following the original gauge configuration
indices of $S^Y$ before any cross-ensemble concatenation is performed.
This step does not simply merge subset blocks; rather, it guarantees
that both the cardinality and ordering of configurations are preserved
exactly.
After this reconstruction, the per-ensemble datasets are concatenated
across ensembles with different $\kappa$ values, forming a unified
configuration stream for the multi-ensemble reweighting while
maintaining the internal ordering within every ensemble.
This index-aligned construction is not unique; in principle, one could
inject ML-predicted subsets earlier in the workflow.
However, the stage at which predictions enter the pipeline has direct
consequences for how prediction-induced inaccuracies propagate through
the subsequent reweighting and (in particular) through the
bias-correction logic.

In these alternative constructions, prediction-induced inaccuracies
from the unlabeled sector can propagate structurally into the
components used for bias correction and labeled-set evaluation after
reweighting.
To mitigate this effect, we construct index-aligned datasets at full
cardinality for $\mathcal{S}_1$ and $\mathcal{S}_2$, and restrict
$\mathcal{S}_3$ and $\mathcal{S}_4$ to labeled-sector configurations,
thereby preserving the intended role of bias correction.
A detailed comparison with alternative constructions is provided in
Appendix~\ref{sec:app-multi-rw}.

Since the cumulants are defined in terms of the ensemble averages
$\langle Q_i \rangle$ [Eqs.~\eqref{eq:cond}--\eqref{eq:C_4}], the
reweighted observables are evaluated at the level of $Q_i(\kappa_r)$.
Bootstrap resampling is performed using identical indices for
$w(\kappa_r)$ and $[w\,Q_i](\kappa_r)$, and the resulting replicas are
propagated pairwise through the cumulant construction.

\begin{figure}[tb]
  \includegraphics[width=\linewidth]{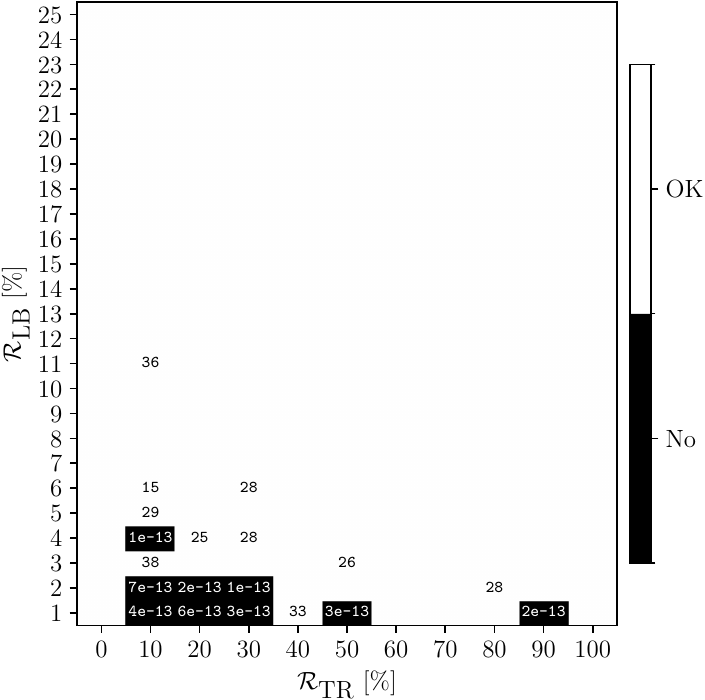}
  \caption{Heatmap of Newton--Raphson convergence for computing the
    free-energy offsets using the $\mathcal{S}_2$ dataset
    (Eq.~\eqref{eq:EA-2}) under the \ref{it:plaquette-feature}
    approach.
    White cells indicate successful convergence before reaching the
    maximum number of iterations; black cells indicate cases that
    failed to find a root within the maximum iterations.
    Numbers inside black cells denote the final residual norm at the
    last iteration.
    White cells are annotated with a number only when convergence
    required more than $10$ iterations.}
  \label{fig:newton-heatmap}
\end{figure}

We now proceed to the practical evaluation of the free-energy offsets,
which are the essential inputs for constructing the reweighting
factors $w_{a;n}(\kappa_r)$ (see Appendix~\ref{sec:app-multi-rw},
steps~\ref{it:R-3}--\ref{it:R-5}).
We compute the free-energy offsets by solving the associated nonlinear
system with \texttt{NLsolve.jl} \cite{NLSolve:2020zen}.
The maximum number of iterations is set to $1000$, and the
function-norm tolerance is set to $10^{-13}$.
With this setup, we solve the Newton--Raphson problem for each set
defined in Eqs.~\eqref{eq:EA-1}--\eqref{eq:EA-4}. Unlike the others,
the solution for $\mathcal{S}_2$ (Eq.~\eqref{eq:EA-2}) failed to find
a root under our prescribed maximum-iteration and tolerance settings
in certain low-$\mathcal{R}_{\text{LB}}$ and
low-$\mathcal{R}_{\text{TR}}$ regions.

As illustrated in Fig.~\ref{fig:newton-heatmap}, the Newton–Raphson
method exhibits a fairly uniform convergence pattern across the
majority of cells, with the number of iterations not exceeding ten in
most cases.
Instances of non-convergence appear only in the very
low–$\mathcal{R}_{\text{LB}}$ region—specifically,
$\mathcal{R}_{\text{LB}}=1\%$ with
$\mathcal{R}_{\text{TR}}=10,\,20,\,30,\,50,\,90\%$,
$\mathcal{R}_{\text{LB}}=2\%$ with
$\mathcal{R}_{\text{TR}}=10,\,20,\,30\%$, and a single case at
$\mathcal{R}_{\text{LB}}=4\%$, $\mathcal{R}_{\text{TR}}=10\%$.
In these cases, the final residual norms remain on the order of
$10^{-13}$ but slightly above the tolerance, and are therefore marked
as non-convergent.
A small set of white cells required $>10$ iterations, appearing
sporadically in the heatmap—predominantly in the lower-left
region—whereas all remaining cells converged within at most $10$
iterations.

Note that for a fixed $\mathcal{R}_{\text{LB}}$, the labeled-set
component of $\mathcal{S}_2$ is, by construction, always the same:
$S^Y_{\text{LB}} = S^Y_{\text{TR}} \cup S^Y_{\text{BC}}$.
Despite this use of identical original data along a given
$\mathcal{R}_{\text{LB}}$ line, the Newton--Raphson convergence is not
uniform at fixed $\mathcal{R}_{\text{LB}}$: as
$\mathcal{R}_{\text{LB}}$ increases, the problematic corner is
increasingly confined toward smaller $\mathcal{R}_{\text{TR}}$.

This variation in convergence \emph{along the same}
$\mathcal{R}_{\text{LB}}$ is clearly driven by the predicted
unlabeled-set component $S^P_{\text{UL}}$.
In the trace estimation stage [Sec.~\ref{sec:trace}], a fraction
$\mathcal{R}_{\text{TR}}$ of configurations is selected as the
training set; a model is trained on that set, and the traces for the
remaining configurations are then predicted to form $S^P_{\text{UL}}$.
At this pre-bootstrap point (prior to bias correction), these
predictions are used as is, and their influence manifests as the
observed differences in Newton--Raphson convergence along constant
$\mathcal{R}_{\text{LB}}$.

In the ensuing multi-ensemble reweighting analysis, we explicitly flag
all cases where the Newton--Raphson evaluation of the free-energy
offsets did not converge.
Given that the residual norms at the non-convergent points exceeded
the tolerance yet remained on the order of $10^{-13}$, and that our
immediate aim is to characterize overall trends, we used the final
Newton--Raphson iterate as a \emph{last-iterate approximation} and
carried out the multi-ensemble reweighting on that basis while
retaining the non-convergence labels in all figures.

To reiterate, the convergence issue described above is specific to the
\ref{it:plaquette-feature} approach.
By contrast, the \ref{it:trace-feature} approach does not exhibit this
behavior under otherwise identical conditions, because the cumulants
are dominated by $\Tr \, M^{-1}$ and that quantity is taken directly
from the original measurements; consequently, its native variability
is faithfully preserved throughout the computation.

Having discussed the convergence behavior of the Newton--Raphson
solver for determining the free-energy offsets in the
\ref{it:plaquette-feature} approach, we first present the results of
the multi-ensemble reweighting analysis obtained within this approach,
and subsequently discuss those from the \ref{it:trace-feature}
approach.

The most direct and intuitive measure of the quality of the ML
estimation for the multi-ensemble reweighting is the comparison
between the cumulants evaluated at the transition point $\kappa_t$
using the primary estimator $\mathcal{P}1$ and those obtained from the
original data, as specified in \ref{it:R-12} of the multi-ensemble
reweighting procedure.

\begin{figure}[tb]
  \includegraphics[width=0.97\linewidth]{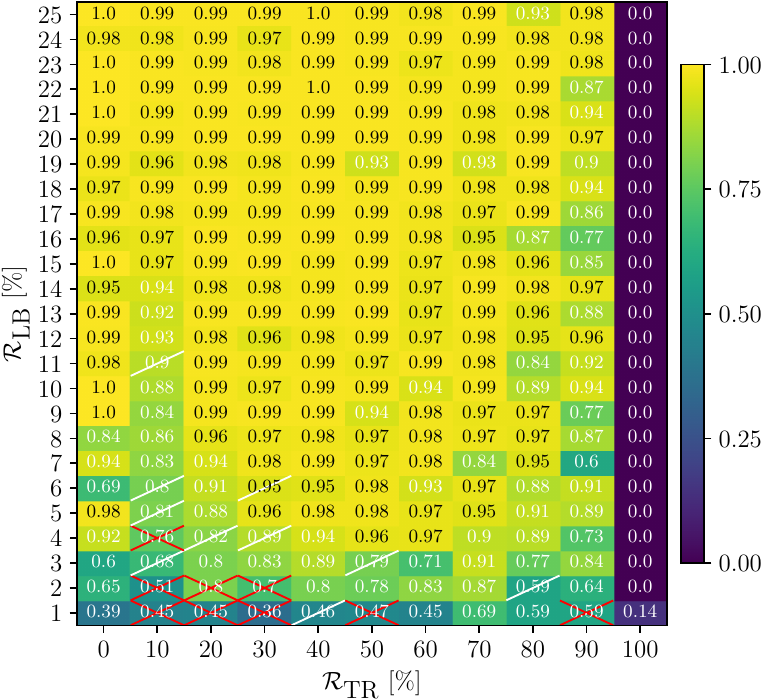}
  \caption{Bhattacharyya coefficient $C_\text{B}$ map for the primary
    estimator $\mathcal{P}1$ in the kurtosis estimation, performed
    using plaquette and rectangle as the input features within the
    \ref{it:plaquette-feature} approach, showing the results for
    $K(\kappa_t)$.
    Red crosses indicate cells corresponding to black cells in
    Fig.~\ref{fig:newton-heatmap}, where the Newton--Raphson solver
    failed to converge within the maximum iteration count.
    White diagonal marks denote cases that converged but required more
    than $10$ iterations.}
  \label{fig:kurt-kurt-heatmap-ML2-LBP-1-25}
\end{figure}

\begin{figure*}[tb]
  \includegraphics[width=0.97\linewidth]{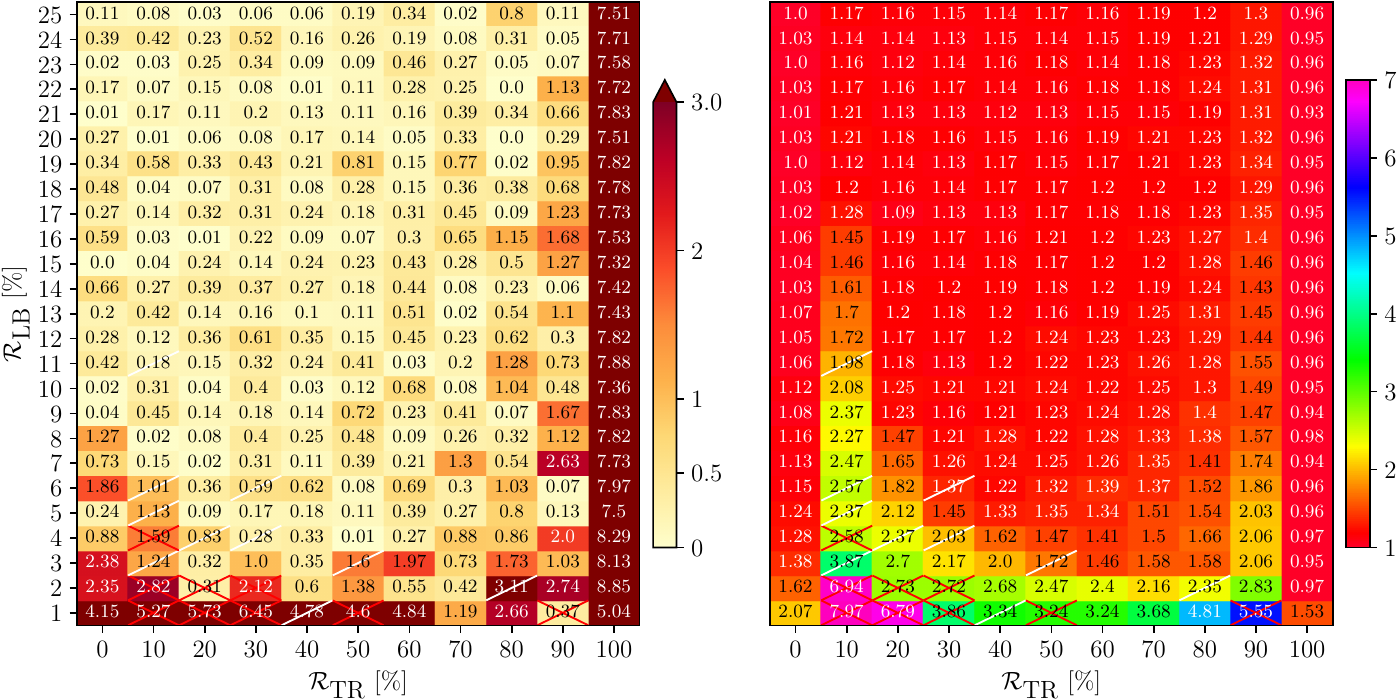}
  \caption{ Heatmap showing the $\mathcal{P}1$ results of
    \ref{it:EC-1} and \ref{it:EC-2} evaluations for the kurtosis
    estimation based on the \ref{it:plaquette-feature} approach,
    showing the results for $K(\kappa_t)$.
    Within the panel, the left half represents the \ref{it:EC-1}
    results, while the right half corresponds to \ref{it:EC-2}.
    In the \ref{it:EC-1} maps, lighter shades indicate smaller
    normalized mean separations $x$ (defined in Eq.~\eqref{eq:x-1}),
    corresponding to better agreement between the original and
    predicted means, whereas darker shades represent larger $x$
    values, signaling poorer overlap.
    The kurtosis is computed using $\Tr\,M^{-n}$ predicted from
    plaquette and rectangle as the input feature.}
  \label{fig:kurt-kurt-heatmap-ML2-LBP-1-25-EC1-EC2}
\end{figure*}

As an illustrative example, we present
Fig.~\ref{fig:kurt-kurt-heatmap-ML2-LBP-1-25}.
This figure shows the results of the multi-ensemble reweighting
obtained with the \ref{it:plaquette-feature} approach, using all
ensembles listed in Table~\ref{tab:ensemble-1}.
Following the interpolation procedure described in \ref{it:R-12}, the
transition point $\kappa_t$ was first estimated from the reweighted
kurtosis curves obtained via both the conventional method using the
original data and ML estimation.
Using these transition points, the corresponding kurtosis values
$K(\kappa_t)$ were then evaluated for the original data and ML
estimation.
The two sets of $K(\kappa_t)$ values are compared by computing the
Bhattacharyya coefficient $C_\text{B}$ according to \ref{it:EC-3},
which quantifies the consistency between the original-data and ML
estimations at the $\kappa_t$.

In this figure, red cross marks indicate the cells corresponding to
the black cells in Fig.~\ref{fig:newton-heatmap}, where the
Newton--Raphson solver for the free-energy offsets did not converge to
the prescribed tolerance within the maximum number of iterations.
Such results are therefore not regarded as valid solutions according
to our convergence criterion; nevertheless, as discussed earlier, they
were included in the subsequent computation of the reweighting
factors, and special care must be taken when interpreting these cells.
White diagonal marks correspond to the white cells with annotated
numbers in Fig.~\ref{fig:newton-heatmap}, representing cases where
convergence was achieved but required more than $10$ iterations.

These red-cross cells are predominantly concentrated in the region
where both $\mathcal{R}_{\text{LB}}$ and $\mathcal{R}_{\text{TR}}$ are
small, and within this region, the Bhattacharyya coefficient
$C_\text{B}$ scores are also noticeably low.
A general tendency can be observed that, in the region characterized
by low $\mathcal{R}_{\text{LB}}$ and low $\mathcal{R}_{\text{TR}}$,
the red-cross and white-diagonal cells are densely distributed, and
even the neighboring cells without such markers tend to exhibit low
$C_\text{B}$ values.

This behavior is consistent with inaccuracies in the predicted dataset
$S^P_{\text{UL}}$ propagating into the reweighting factors.
Larger prediction deviations can in turn make the Newton--Raphson
determination of the free-energy offsets more difficult to converge.
Even when the solver converges owing to the increasing contribution
from the original data with larger $\mathcal{R}_{\text{LB}}$ or
$\mathcal{R}_{\text{TR}}$, residual inaccuracies originating from
$S^P_{\text{UL}}$ may still result in relatively low $C_\text{B}$
scores.
In Fig.~\ref{fig:kurt-kurt-heatmap-ML2-LBP-1-25}, this trend appears
to be largely mitigated once $\mathcal{R}_{\text{LB}}$ exceeds
approximately $20\,\%$.

In line with the trends already observed in Secs.~\ref{sec:trace} and
\ref{sec:cumulant-single} for the evaluation of the ML estimations,
regions with small $\mathcal{R}_{\text{LB}}$ and large
$\mathcal{R}_{\text{TR}}$ (equivalently, small
$\mathcal{R}_{\text{BC}} = 1 - \mathcal{R}_{\text{TR}}$) are once
again found to yield relatively lower $C_\text{B}$ scores.

However, in the column corresponding to $\mathcal{R}_{\text{TR}} =
90\,\%$, the $C_\text{B}$ scores become stabilized only when
$\mathcal{R}_{\text{LB}}$ exceeds approximately $23\,\%$, showing a
trend quite similar to that observed for the $\mathcal{R}_{\text{TR}}
= 10\,\%$ column.
This indicates that the quality of the ML estimation tends to be
relatively low either when the training set fraction is extremely
small or, conversely, when the bias correction set fraction is very
small.

Another notable observation is that, in the $\mathcal{R}_{\text{TR}} =
100\,\%$ column—where the entire labeled set is used solely for model
training and no bias correction is performed—the $C_\text{B}$ scores
are essentially zero.
To investigate the cause of this behavior, we performed an analysis
similar to that in Sec.~\ref{sec:cumulant-single}, separating the
evaluation according to the criteria defined in \ref{it:EC-1} and
\ref{it:EC-2}.
The results are presented in
Fig.~\ref{fig:kurt-kurt-heatmap-ML2-LBP-1-25-EC1-EC2}.
The figure reports the primary estimator $\mathcal{P}1$.
The left panel of
Fig.~\ref{fig:kurt-kurt-heatmap-ML2-LBP-1-25-EC1-EC2} shows that $x =
{\lvert Y_{\mathcal{P}1} - Y_{\text{OG}} \rvert} /
{\sigma_{\text{OG}}}$ takes values around $7$--$8$ on average.  A
large mean separation is therefore the dominant diagnostic feature in
these cells.
This indicates that the mean of the $\mathcal{P}1$ results deviates
from that of the original (conventional) data by as much as seven to
eight times the statistical error of the original results.

The (relatively) poor $C_\text{B}$ scores observed in the
$\mathcal{R}_{\text{TR}} = 100\,\%$ column were already noted to some
extent in Sec.~\ref{sec:cumulant-single}, but the discrepancy becomes
even more pronounced in the present multi-ensemble reweighting and
interpolation results for $K(\kappa_t)$.
It is likely that this relatively large deviation has appeared more
prominently due to the particular characteristics of the dataset used
here.
Within the fixed dataset analyzed here, these cells therefore provide
a descriptive indication that omitting bias correction can amplify
deviations during the nonlinear reweighting analysis; they do not
establish a dataset-independent effect size.

\begin{figure}[tb]
  \includegraphics[width=0.97\linewidth]{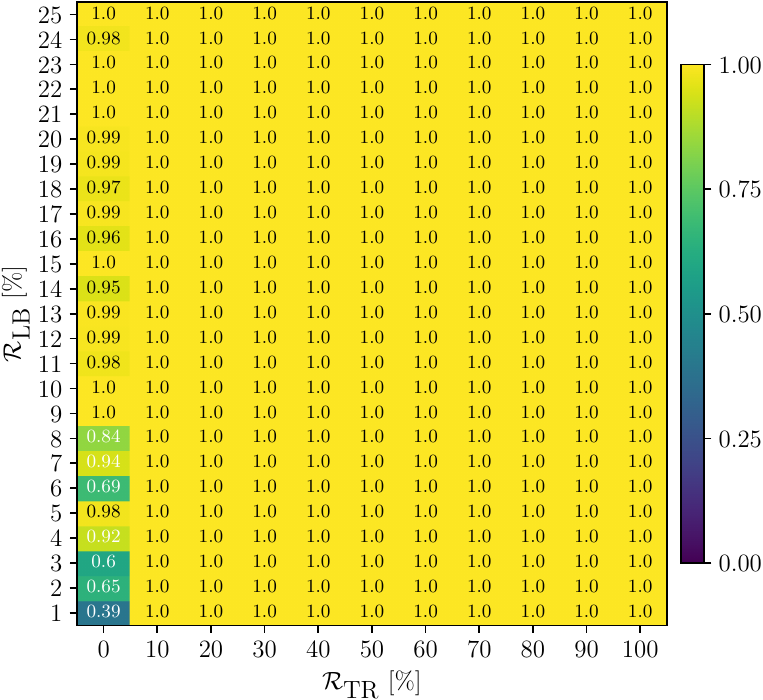}
  \caption{A Bhattacharyya coefficient $C_\text{B}$ map for
    $\mathcal{P}1$ in the kurtosis estimation, performed using
    $\Tr\,M^{-1}$ from the original data and $\Tr\,M^{-n}$ ($n=2,3,4$)
    predicted from $\Tr\,M^{-1}$ as the input feature within the
    \ref{it:trace-feature} approach, showing the results for
    $K(\kappa_t)$.}
  \label{fig:kurt-kurt-heatmap-ML1-LBP-1-25}
\end{figure}
%
Finally, let us discuss the results of the multi-ensemble reweighting
obtained using the \ref{it:trace-feature} approach.
The corresponding results are shown in
Fig.~\ref{fig:kurt-kurt-heatmap-ML1-LBP-1-25}. As can be immediately
seen from the figure, the $\mathcal{P}1$ results already achieve
$C_\text{B}$ scores that are essentially $1.0$ across the entire
scanned region; therefore, only the $\mathcal{P}1$ results are
presented here for conciseness.
The only noticeable feature is the lower $C_\text{B}$ scores in the
low-$\mathcal{R}_{\text{LB}}$ region of the $\mathcal{R}_{\text{TR}} =
0\,\%$ column.
This occurs simply because the dataset corresponding to
$\mathcal{R}_{\text{TR}} = 0\,\%$ assumes no ML estimation procedure
was applied at all — only the labeled set was available, and both the
mean and the error were computed solely from that set.
In such a case, the statistical precision is inevitably much poorer
compared with that of the full original dataset, and the lower
$C_\text{B}$ values are merely a natural consequence of this
limitation.

\section{Conclusion}
\label{sec:conc}

We have conducted a retrospective case study of bias-corrected
machine-learning estimation for chiral-condensate cumulants in a fixed
lattice QCD dataset.
The primary analysis uses the estimator $\mathcal{P}1$ and compares it
with the conventional full-data result at three successive stages:
direct estimation of $\Tr\,M^{-n}$, single-ensemble cumulants, and
multi-ensemble reweighting.
The alternative weighted estimator $\mathcal{P}2$ and its uncertainty
diagnostics are reported separately in Appendix~\ref{sec:app-p2}.

For the \ref{it:trace-feature} approach, the ML-based estimates show
close agreement with the full-data estimates under the adopted
criteria throughout the parameter scan considered here.
Much of this agreement can be traced to a shared dominant component:
the complete original $\Tr\,M^{-1}$ data are used as the feature,
while the higher-order traces are predicted.
The result should therefore be understood as a property of this
estimator and this fixed dataset, not as independent replication of
the full calculation.

Under a nominal solve-count model, the corresponding Dirac-inversion
cost is
\begin{equation}
  \frac{100+1+1+1}{400} \simeq 25.75\% .
\end{equation}
Thus, the retrospective analysis of this fixed dataset indicates the
possibility of reducing the solver cost to approximately $25.75\%$ of
the conventional calculation.
This is a cost projection, not an end-to-end timing benchmark: it
assumes comparable costs for the successive inversions and excludes
model training, data handling, and downstream analysis overhead.
Nor does the observed agreement by itself establish equivalent
precision on new ensembles.

For the \ref{it:plaquette-feature} approach, the fixed-dataset results
show a less uniform pattern.
Agreement generally improves as the labeled fraction increases, and
the explored scan becomes relatively stable around
$\mathcal{R}_{\mathrm{LB}}\gtrsim20\,\%$ for several of the reported
observables.
This value is a descriptive feature of the present scan rather than a
transferable threshold or a second cost estimate.
The results without bias correction also show larger deviations after
the nonlinear cumulant and reweighting steps, suggesting that small
residual biases can be amplified in downstream observables within this
dataset.

These conclusions are conditional on the deterministic partition used
here.
The reported block-bootstrap uncertainties depend on the empirically
selected block length and do not include variation from alternative
train--correction splits or model retraining.
In addition, the Gaussian Bhattacharyya coefficient is a descriptive
overlap measure, not a test of statistical equivalence, and the ML and
full-data estimates are correlated because they share observations.
Subject to these qualifications, the study provides evidence of
feasibility and a concrete solve-count target for future prospective
validation, without requiring that target to be interpreted as an
already demonstrated end-to-end speedup.

\acknowledgments

B.~J.~C. would like to thank Takayuki Sumimoto for his early
contributions and dedication to the initial stage of this work.
He also thanks Ho Hsiao for fruitful discussions.
The work of A.~T. was partially supported by JSPS KAKENHI Grants
No.~20K14479, No.~22H05111, No.~22K03539 and JST BOOST, Japan Grant
No.~JPMJBY24F1.
B.~J.~C., A.~T. and H.~O. were partially supported by JSPS KAKENHI
Grant No.~22H05112.
B.~J.~C. and part of this work were supported by MEXT as ``Program for
Promoting Researches on the Supercomputer Fugaku'' (Grant Number
JPMXP1020230411, JPMXP1020230409).

\appendix

\section{Cumulants of chiral condensate}
\label{sec:cumulants-app}
%
The partition function for the Wilson–Clover quark action is given by
\begin{align}
  Z &= \int \mathcal{D}U \; e^{-S_g} \, \left( \det M
  \right)^{N_{\text{f}}} \,, \label{eq:partition-1}
\end{align}
where $N_{\text{f}}$ denotes the number of quark flavors.
In this study, we explicitly consider the case of $N_{\text f}=4$
degenerate flavors and therefore keep $N_{\text f}$ explicit for
clarity.

The expectation value of the chiral condensate, $\Sigma = \langle
\bar{\psi} \psi \rangle$, is given by
\begin{align}
  \Sigma &= \frac{1}{V} \, \frac{\partial}{\partial m} \ln Z \nonumber
  \\
  &= \frac{1}{V} \, \frac{1}{Z} \int \mathcal{D}U \; e^{- S_g +
    N_{\text{f}} \ln \det M} \; \frac{\partial \ln \left(\det M
    \right)^{N_{\text{f}}}}{\partial m} \,. \label{eq:Sigma-1}
\end{align}
At this point we need the $m$–derivative of $\ln\!\left(\det
M\right)^{N_{\text f}}$.
We write the Dirac operator as $M(m)=m\,\mathbb{1}+D_f$, where $D_f$
is $m$-independent, so that
\begin{equation}
  \frac{\partial}{\partial m} M = \mathbb{1} \,. \label{eq:dM-unity-1}
\end{equation}
Therefore,
\begin{align}
  \frac{\partial}{\partial m}\ln\left(\det M\right)^{N_{\text{f}}}
  &=N_{\text{f}}\;\frac{\partial}{\partial m}\ln\det M\\
  &=N_{\text{f}}\;\frac{\partial}{\partial m}\Tr\ln M\\
  &=N_{\text{f}}\;\Tr\left(\frac{\partial}{\partial m}\ln M\right)\\
  &=N_{\text{f}}\;\Tr\left(M^{-1}\frac{\partial}{\partial m}M\right)\\
  &=N_{\text{f}}\;\Tr\,M^{-1}\,. \label{eq:ln-detM-m-1}
\end{align}
Here we note that, for any matrix $M$ depending smoothly on a real
parameter $m$, the trace operation is simply the sum of its diagonal
entries:
\begin{align}
  \Tr \, M &= \sum_{i}M_{ii} \,.
\end{align}
Differentiation with respect to $m$ therefore acts element-wise,
giving
\begin{align}
  \frac{\partial}{\partial m}\Tr\,M &= \sum_{i}\frac{\partial
    M_{ii}}{\partial m}=\Tr \left(\frac{\partial}{\partial m} M
  \right)\,.
\end{align}
Hence, for any differentiable matrix function $F(M(m))$ that is
analytic in $M$ (\textit{e.g.}~a polynomial or power-series function
such as $M^{-k}$, $\exp M$, or $\log M$), differentiation with respect
to $m$ and the trace operation can be interchanged:
\begin{align}
  \frac{\partial}{\partial m}\Tr\,F(M)=\Tr
  \left(\frac{\partial}{\partial m} F\left(M\right) \right)
  \,. \label{eq:dm-Tr-commute-1}
\end{align} 
With the usage of Eq.~\eqref{eq:dM-unity-1},
Eq.~\eqref{eq:dm-Tr-commute-1} proves Eq.~\eqref{eq:ln-detM-m-1}
rigorously.
Substituting Eq.~\eqref{eq:ln-detM-m-1} into Eq.~\eqref{eq:Sigma-1}
then yields
\begin{align}
  \Sigma &= \frac{ \left\langle N_{\text{f}} \, \Tr\,M^{-1}
    \right\rangle}{V} \,.
\end{align}

For the calculation of the $k$-th moment of the chiral condensate, we
require the $k$-th derivative of $\left(\det M\right)^{N_{\text{f}}}$
with respect to $m$.
Using Eq.~\eqref{eq:ln-detM-m-1}, the first derivative follows
directly:
\begin{align}
  \frac{\partial}{\partial m}\left(\det M\right)^{N_{\text{f}}} &=
  \frac{\partial}{\partial m}\exp\left[N_{\text{f}}\;\ln\det M\right]
  \nonumber \\
  &= \left(\det
  M\right)^{N_{\text{f}}}\;N_{\text{f}}\;\frac{\partial}{\partial
    m}\ln\det M \nonumber \\
  &= \left(\det M\right)^{N_{\text{f}}} \; N_{\text{f}}\;\Tr\,M^{-1}
  \,.
    \label{eq:ddetM-dm-1}
\end{align}
For conciseness, we adopt Fa{\`a} di Bruno's formula
\cite{FaaDiBruno:1855aa,FaaDiBruno:1857bb} --- a compact
generalization of the chain rule to higher derivatives:
\begin{align}
  \frac{\partial^{k}}{\partial m^{k}}(\det M)^{N_{f}}=(\det M)^{N_{f}}
  \; Q_{k} \,, \label{eq:bruno-bell-1}
\end{align}
where $Q_k$ denotes, in physical terms, the operator corresponding to
the $k$-th moment, and, in mathematical terms, the $k$-th complete
(exponential) Bell polynomial \cite{Bell:1934aa},
\begin{align}
  Q_{1}&=W_{1} \,, \label{eq:B_1} \\
  Q_{2}&=W_{1}^{2}+W_{2} \,, \label{eq:B_2} \\
  Q_{3}&=W_{1}^{3}+3W_{1}W_{2}+W_{3} \,, \label{eq:B_3} \\
  Q_{4}&=W_{1}^{4}+6\,W_{1}^{2}W_{2}+3\,W_{2}^{2}+4\,W_{1}W_{3}+W_{4}
  \,, \label{eq:B_4}
\end{align}
and where $W_i$ is defined as
\begin{align}
  W_{k} & \equiv \frac{\partial^{k}}{\partial m^{k}} \ln \left(\det
  M\right)^{N_{\text{f}}} \nonumber \\
  & = N_{\text{f}} \; \left(-1\right)^{k-1} \; \left(k-1\right)! \;
  \Tr \, M^{-k} \,. \label{eq:dklndetM-dkm-1}
\end{align}
As noted in passing in the derivation of Eq.~\eqref{eq:ddetM-dm-1}, we
write $\left(\det M\right)^{N_{\text{f}}} =
\exp\!\left[N_{\text{f}}\;\ln\det M\right]$, which allows us to employ
Eqs.~\eqref{eq:B_1}--\eqref{eq:B_4} in Eq.~\eqref{eq:bruno-bell-1}.
From Eqs.~\eqref{eq:B_1}--\eqref{eq:dklndetM-dkm-1}, the explicit
forms follow, as listed in Eqs.~\eqref{eq:Q_1}--\eqref{eq:Q_4}.

Here we give a brief proof of Eq.~\eqref{eq:dklndetM-dkm-1} by
mathematical induction.
The case $k=1$ follows from Eq.~\eqref{eq:ddetM-dm-1}.
Assume it holds for some $k \ge 1$.
Using $M^{-1} \; M = \mathbb{1}$ and
\begin{equation}
  \frac{\partial}{\partial m} M^{-1} = -
  M^{-1}\!\left(\frac{\partial}{\partial m} M\right) M^{-1} = -
  M^{-2},
\end{equation}
together with the product rule, we obtain
\begin{align}
  \frac{\partial}{\partial m} \, \Tr \, M^{-k} &= - k \Tr\,M^{-(k+1)}
  \,.
  \label{eq:deriv-rule-base}
\end{align}
For completeness, we verify Eq.~\eqref{eq:deriv-rule-base}.
The $m$-derivative of $M^{-k}$ is
\begin{align}
  \frac{\partial}{\partial m} M^{-k} &= \sum_{r=0}^{k-1}
  \left(M^{-1}\right)^{r} \left(\frac{\partial}{\partial m}
  M^{-1}\right) \left(M^{-1}\right)^{k-1-r} \nonumber \\
  &= - \sum_{r=0}^{k-1} \left(M^{-1}\right)^{r} M^{-2}
  \left(M^{-1}\right)^{k-1-r} \nonumber \\
  &= - \sum_{r=0}^{k-1} M^{-(k+1)} = - k M^{-(k+1)} \,.
  \label{eq:dM-dm-1}
\end{align}
Taking the trace of Eq.~\eqref{eq:dM-dm-1} yields
\begin{align}
  \Tr \left(\frac{\partial}{\partial m} M^{-k}\right)
  &= \frac{\partial}{\partial m} \, \Tr \, M^{-k} \nonumber \\
  &= - k \Tr\,M^{-(k+1)} \,,
  \label{eq:dTrM-dm-1}
\end{align}
which is Eq.~\eqref{eq:deriv-rule-base}.
Differentiating Eq.~\eqref{eq:dklndetM-dkm-1} once more and using
Eq.~\eqref{eq:dTrM-dm-1}, we obtain
\begin{align}
  W_{k+1} &= N_{\text{f}} \; (-1)^{k-1} \; (k-1)! \; \Tr
  \left(\frac{\partial}{\partial m} M^{-k}\right) \nonumber \\
  &= N_{\text{f}} \; (-1)^{k} \; k! \; \Tr \, M^{-(k+1)} \,,
\end{align}
which completes the induction.

Using Eqs.~\eqref{eq:partition-1} and \eqref{eq:bruno-bell-1}, the
expectation value of the $j$-th moment operator (hereafter, the $j$-th
moment) is
\begin{align}
  \left\langle Q_j \right\rangle & = \frac{Z^{(j)}}{Z} = \frac{1}{Z}
  \frac{\partial^j Z}{\partial m^j} \,. \label{eq:moment-ev-1}
\end{align}

We now define $C_j$ as the $j$-th cumulant. In close analogy to the
moment case, we again apply Fa\`a di Bruno’s formula; however, the
right-hand side now involves the \emph{partial} Bell polynomials:
\begin{equation}
  C_{j} = \frac{\partial^{j}}{\partial m^{j}}\ln Z = \sum_{k=1}^{j}
  (-1)^{k-1}(k-1)! \; \frac{B_{j,k}}{Z^{k}}
  \,, \label{eq:cumulant-def-1}
\end{equation}
where
\begin{align}
  B_{j,1} &= Z^{(j)} \qquad\quad\;\;\; (j = 1,2,3,4)
  \,, \label{eq:B-k1-1} \\
  B_{j,j} &= \left(Z^{(1)}\right)^{j} \qquad (j = 2,3,4)
  \,, \label{eq:B-kk-1} \\
  B_{3,2} &= 3 \; Z^{(2)} \; Z^{(1)} \,, \label{eq:B-32-1} \\
  B_{4,2} &= 4 \; Z^{(3)} \; Z^{(1)} + 3 \,
  \left(Z^{(2)}\right)^{2}\,, \label{eq:B-42-1} \\
  B_{4,3} &= 6 \; Z^{(2)} \, \left(Z^{(1)}\right)^{2}
  \,, \label{eq:B-43-1}
\end{align}
and $Z^{(j)}$ is defined in Eq.~\eqref{eq:moment-ev-1}.
Using Eqs.~\eqref{eq:cumulant-def-1}--\eqref{eq:B-43-1}, we obtain
\begin{align}
  C_{1} &= \frac{Z^{(1)}}{Z} \,, \\[0.25em]
  C_{2} &= \frac{Z^{\left(2\right)}}{Z} -
  \left(\frac{Z^{\left(1\right)}}{Z}\right)^{2} \,, \\[0.25em]
  C_{3} &= \frac{Z^{\left(3\right)}}{Z} -
  3\left(\frac{Z^{\left(2\right)}}{Z}\right)
  \left(\frac{Z^{\left(1\right)}}{Z}\right) +
  2\left(\frac{Z^{\left(1\right)}}{Z}\right)^{3} \,, \\[0.25em]
  C_{4} &= \frac{Z^{\left(4\right)}}{Z} -
  4\left(\frac{Z^{\left(3\right)}}{Z}\right)
  \left(\frac{Z^{\left(1\right)}}{Z}\right) -
  3\left(\frac{Z^{\left(2\right)}}{Z}\right)^{2} \nonumber \\
  &\hphantom{=} +12\left(\frac{Z^{\left(2\right)}}{Z}\right)
  \left(\frac{Z^{\left(1\right)}}{Z}\right)^{2} -
  6\left(\frac{Z^{\left(1\right)}}{Z}\right)^{4} \,.
\end{align}
or, using Eq.~\eqref{eq:moment-ev-1},
\begin{align}
  C_{1} &= \left\langle Q_1 \right\rangle \,, \\[0.25em]
  C_{2} &= \left\langle Q_2 \right\rangle - \left\langle Q_1
  \right\rangle^{2} \,, \\[0.25em]
  C_{3} &= \left\langle Q_3 \right\rangle-3\left\langle Q_2
  \right\rangle \left\langle Q_1 \right\rangle + 2 \left\langle Q_1
  \right\rangle^{3} \,, \\[0.25em]
  C_{4} &= \left\langle Q_4 \right\rangle - 4 \left\langle Q_3
  \right\rangle \left\langle Q_1 \right\rangle -3 \left\langle Q_2
  \right\rangle^{2} \nonumber \\
  &\hphantom{=} +12 \left\langle Q_2 \right\rangle \left\langle Q_1
  \right\rangle^{2}-6\left\langle Q_1 \right\rangle^{4} \,,
\end{align}
which reproduce Eqs.~\eqref{eq:C_1}--\eqref{eq:C_4}.
The expectation value of the chiral condensate and its susceptibility,
skewness, and kurtosis are then given by
\begin{align}
  \Sigma &= \frac{1}{V}\frac{\partial}{\partial m}\ln
  Z=\frac{C_{1}}{V}\,,\\
  \chi &= \frac{1}{V}\frac{\partial^{2}}{\partial m^{2}}\ln
  Z=\frac{C_{2}}{V}\,,\\
  S &= \frac{1}{{\displaystyle \left(\frac{\partial^{2}}{\partial
        m^{2}}\ln Z\right)^{\frac{3}{2}}}}\frac{\partial^{3}}{\partial
    m^{3}}\ln Z=\frac{C_{3}}{C_{2}^{\frac{3}{2}}}\,,\\
  K &= \frac{1}{{\displaystyle \left(\frac{\partial^{2}}{\partial
        m^{2}}\ln Z\right)^{2}}}\frac{\partial^{4}}{\partial m^{4}}\ln
  Z=\frac{C_{4}}{C_{2}^{2}}\,,
\end{align}
in agreement with Eqs.~\eqref{eq:cond}--\eqref{eq:kurt}, where
$V=N_\text{S}^3\,N_\text{T}$ denotes the lattice volume.

\section{Deterministic Construction of the Partitioning of the Total Dataset $S^Z$}
\label{sec:app-dataset-split}

Here we describe in detail the deterministic construction of the
partitioning scheme used to divide the total dataset $S^Z$.

The full set of configuration indices $\{1,2,\dots,N\}$ is first
divided into a labeled subset $S^Z_{\text{LB}}$ and an unlabeled
subset $S^Z_{\text{UL}}$.
Unlike random sampling commonly adopted in practical machine-learning
applications, this approach follows a deterministic rule based on
uniform-interval sampling with a cyclic shift.
The motivation for this choice is to preserve, as faithfully as
possible, the overall statistical tendency of the total dataset $S^Z$
within each subset.

For a given integer shift $s$, the indices of the labeled set
$S^Z_{\text{LB}}$ are constructed as
\begin{equation}
  i_{\text{LB}}(j) = \left( \left\lfloor
  (j-1)\,\frac{N}{N_{\text{LB}}} + s \right\rfloor \bmod N \right) + 1
  \, ,
\end{equation}
where $j = 1,\dots,N_{\text{LB}}$ and $\lfloor \cdot \rfloor$ denotes
the floor operation.
The modulo operation ensures that the sampling pattern repeats
cyclically across the dataset.

The shift parameter $s$ determines the starting position of the
labeled set within the full sequence.
Because of the periodicity introduced by the modulo operation,
different shift values can in principle generate identical
$S^Z_{\text{LB}}$ configurations.
To avoid such redundancy, all shifts reproducing the same
$S^Z_{\text{LB}}$ are examined and the smallest non-negative one is
chosen as the canonical representative.
Although nonzero shifts ($s \neq 0$) are allowed in the general
formulation, all numerical results presented in this work are obtained
with $s=0$ for simplicity.

The training set $S^Z_{\text{TR}}$ is constructed by uniformly
sampling $N_{\text{TR}}$ elements from the ordered labeled set
$S^Z_{\text{LB}}$.
The corresponding indices are defined as
\begin{equation}
  i_{\text{TR}}(k) = \left\lfloor
  (k-1)\,\frac{N_{\text{LB}}}{N_{\text{TR}}} + 1 \right\rfloor ,
\end{equation}
where $k = 1,\dots,N_{\text{TR}}$.
This ensures an approximately uniform coverage of $S^Z_{\text{LB}}$
while preserving its internal ordering.

The bias-correction set $S^Z_{\text{BC}}$ is defined as the complement
of the training set within the labeled set, such that $N_{\text{TR}} +
N_{\text{BC}} = N_{\text{LB}}$.

Finally, the unlabeled set $S^Z_{\text{UL}}$ is defined as the
complement of $S^Z_{\text{LB}}$ within the total dataset.
By construction, the subsets satisfy
\begin{equation}
  S^Z = S^Z_{\text{LB}} \cup S^Z_{\text{UL}}, \qquad S^Z_{\text{LB}} =
  S^Z_{\text{TR}} \cup S^Z_{\text{BC}}.
\end{equation}

\section{Alternative weighted estimator $\mathcal{P}2$}
\label{sec:app-p2}

\begin{figure*}[tb]
  \centering
  \subfigure[\ref{it:EC-3} diagnostic with \ref{it:trace-feature} approach]{
    \includegraphics[width=0.48\linewidth]{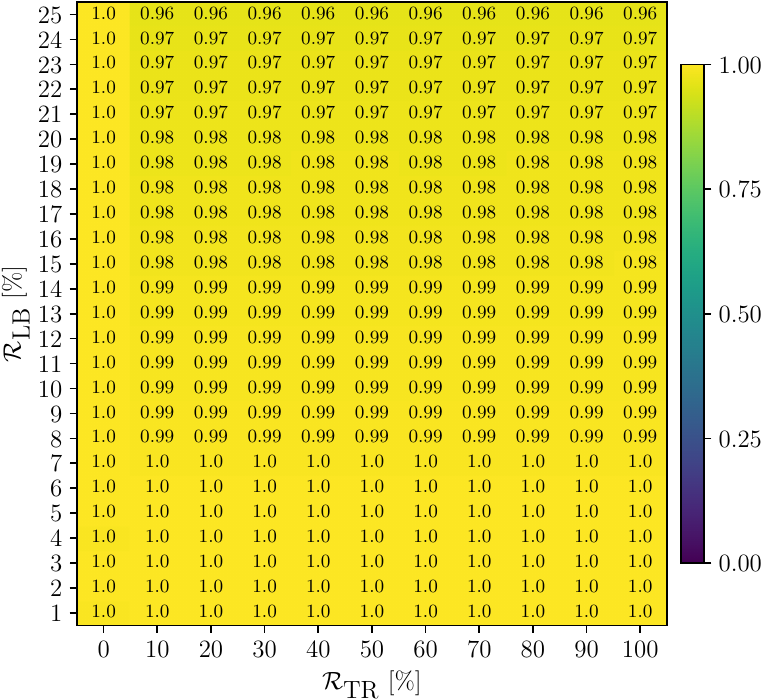}
    \label{fig:subfig:app-p2-trace-1}
  }
  \hfill
  \subfigure[\ref{it:EC-3} diagnostic with \ref{it:plaquette-feature} approach]{
    \includegraphics[width=0.48\linewidth]{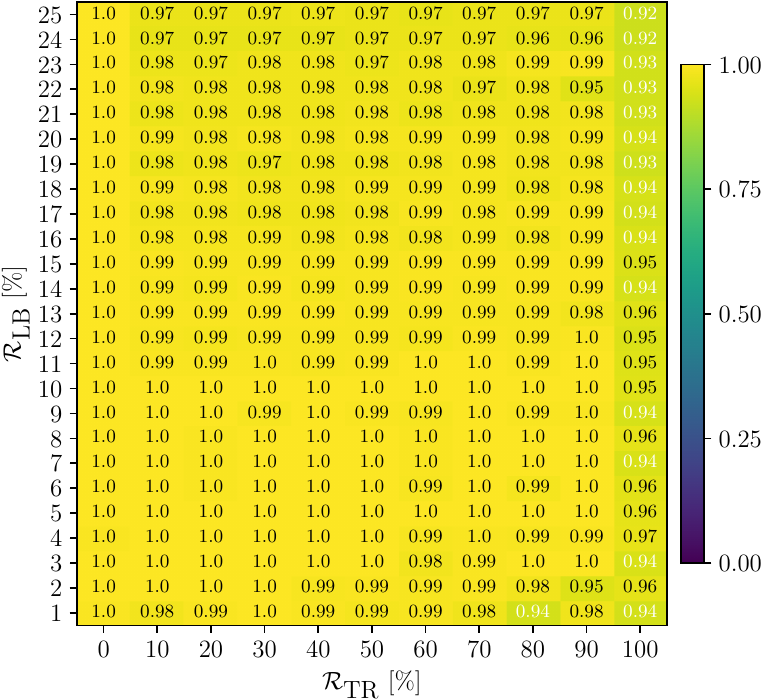}
    \label{fig:subfig:app-p2-trace-2}
  }
  \subfigure[\ref{it:EC-1} and \ref{it:EC-2} diagnostics with \ref{it:plaquette-feature} approach]{
    \includegraphics[width=0.98\linewidth]{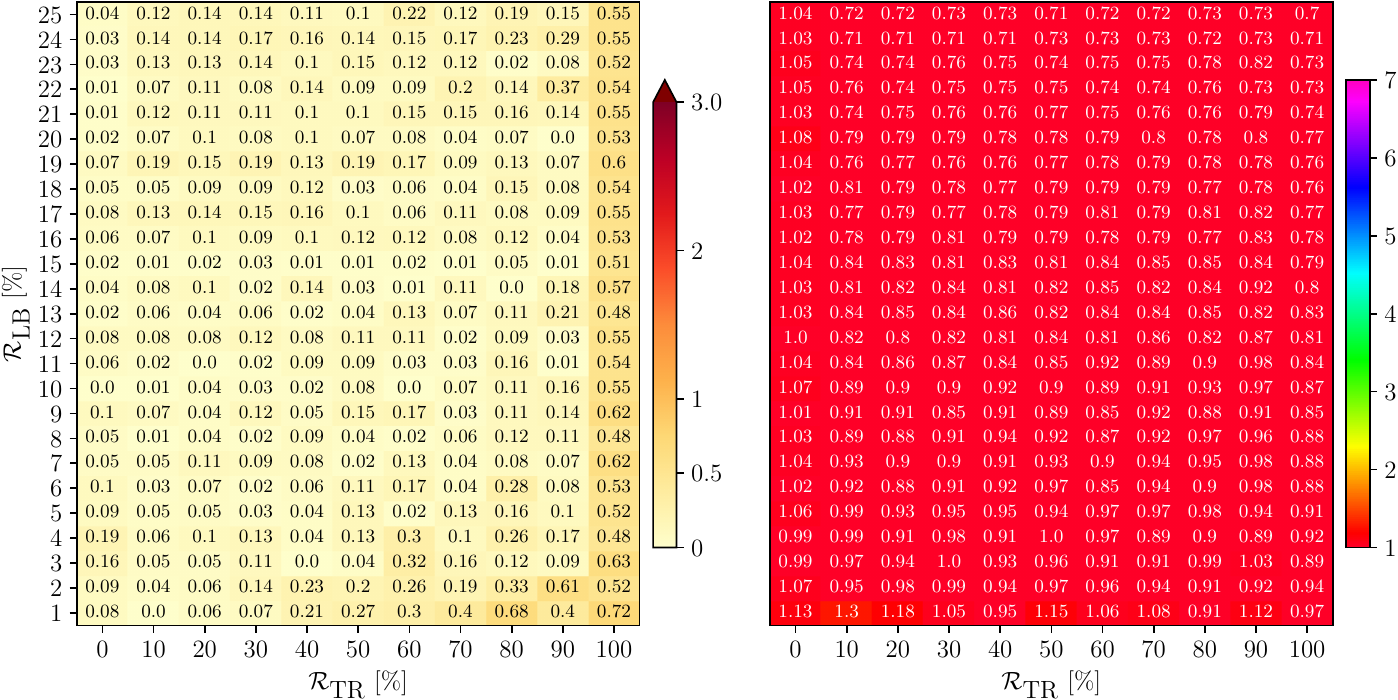}
    \label{fig:subfig:app-p2-trace-3}
  } 
  \caption{Diagnostic maps for the $\mathcal{P}2$ estimate of the
    single-ensemble kurtosis for dataset \texttt{L12T4b1.60k13580}.
    Panels (a) and (b) show the \ref{it:EC-3} (Bhattacharyya
    coefficient $C_B$) diagnostic obtained with the
    \ref{it:trace-feature} and \ref{it:plaquette-feature} approaches,
    respectively.
    Panel (c) shows the corresponding \ref{it:EC-1} and \ref{it:EC-2}
    diagnostics for the \ref{it:plaquette-feature} approach.
    These maps are used to characterize the behavior of $\mathcal{P}2$
    over the scanned labeled and training fractions; they are not
    interpreted as evidence that $\mathcal{P}2$ achieves a genuine
    improvement in statistical precision.}
  \label{fig:app-p2-trace}
\end{figure*}

For completeness, this appendix records the alternative weighted
estimator examined during the analysis.
It is defined by
\begin{align}
  \bar{Y}_{\mathcal{P}2} &=
  \frac{N_{\textrm{UL}}}{N}\,\bar{Y}_{\mathcal{P}1} +
  \frac{N_{\textrm{LB}}}{N}\,\bar{Y}_{(\text{LB})},
  \label{eq:P2-1} \\
  \bar{Y}_{(\text{LB})} &= \frac{1}{N_{\text{LB}}} \sum_{Y_k\in
    S^Y_{\text{LB}}}Y_k, \qquad N=N_{\text{LB}}+N_{\text{UL}} \,
  . \nonumber
\end{align}
A related weighted-average construction was discussed in
Ref.~\cite{Yoon:2018krb}.
In the fixed dataset studied here, $\mathcal{P}2$ sometimes produces a
smaller estimated uncertainty than both $\mathcal{P}1$ and the
full-data reference.
For example, the single-ensemble kurtosis diagnostics shown in
Fig.~\ref{fig:app-p2-trace} exhibit an apparently reduced uncertainty
for $\mathcal{P}2$ in part of the scanned parameter region.
In particular, for the \ref{it:plaquette-feature} approach, the
error-ratio diagnostic decreases from approximately $0.88$ in part of
the scanned region to approximately $0.7$ near
$\mathcal{R}_{\mathrm{LB}}=25\,\%$
[Fig.~\ref{fig:app-p2-trace}\subref{fig:subfig:app-p2-trace-3}].
The corresponding Bhattacharyya coefficient $C_B$ maps for the
\ref{it:trace-feature} and \ref{it:plaquette-feature} approaches are
displayed in
Figs.~\ref{fig:app-p2-trace}\subref{fig:subfig:app-p2-trace-1} and
\ref{fig:app-p2-trace}\subref{fig:subfig:app-p2-trace-2},
respectively.
Because the weighted terms reuse correlated labeled and
bias-correction information, interpreting this reduction would require
a covariance treatment beyond the present retrospective analysis.
We therefore do not interpret it as an improvement in precision and
use $\mathcal{P}1$ as the sole primary estimator in the main text.

\begin{figure*}[tb]
  \centering
  \subfigure[\ref{it:trace-feature} approach,
    \texttt{L12T4b1.60k13585}]{
    \includegraphics[width=0.47\linewidth]{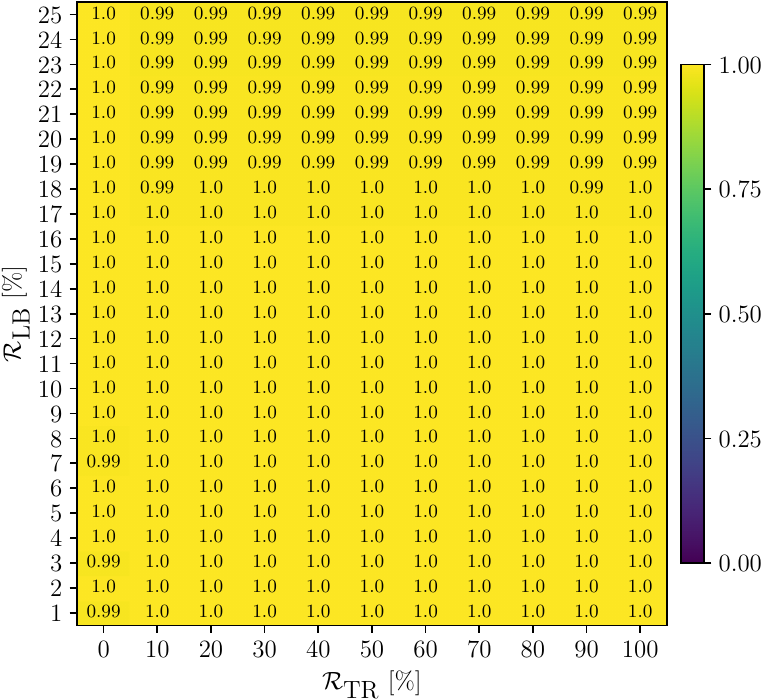}
    \label{fig:subfig:app-p2-cumulants-1}
  }
  \hfill
  \subfigure[\ref{it:plaquette-feature} approach,
    \texttt{L12T4b1.60k13585}]{
    \includegraphics[width=0.47\linewidth]{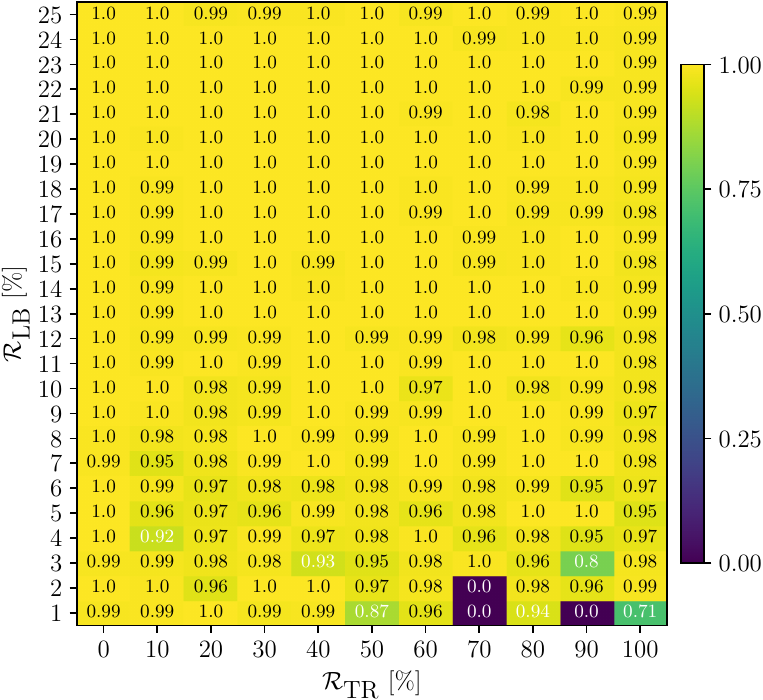}
    \label{fig:subfig:app-p2-cumulants-2}
  }
  \hfill
  \subfigure[\ref{it:plaquette-feature} approach,
    \texttt{L12T4b1.60k13582}]{
    \includegraphics[width=0.47\linewidth]{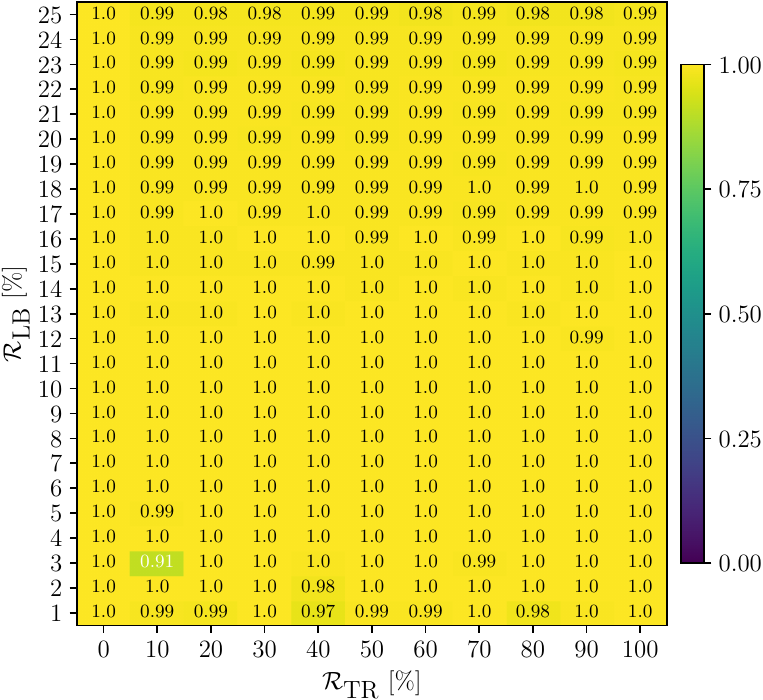}
    \label{fig:subfig:app-p2-cumulants-3}
  }
  \hfill
  \subfigure[Multi-ensemble reweighting with \ref{it:plaquette-feature} approach]{
    \includegraphics[width=0.47\linewidth]{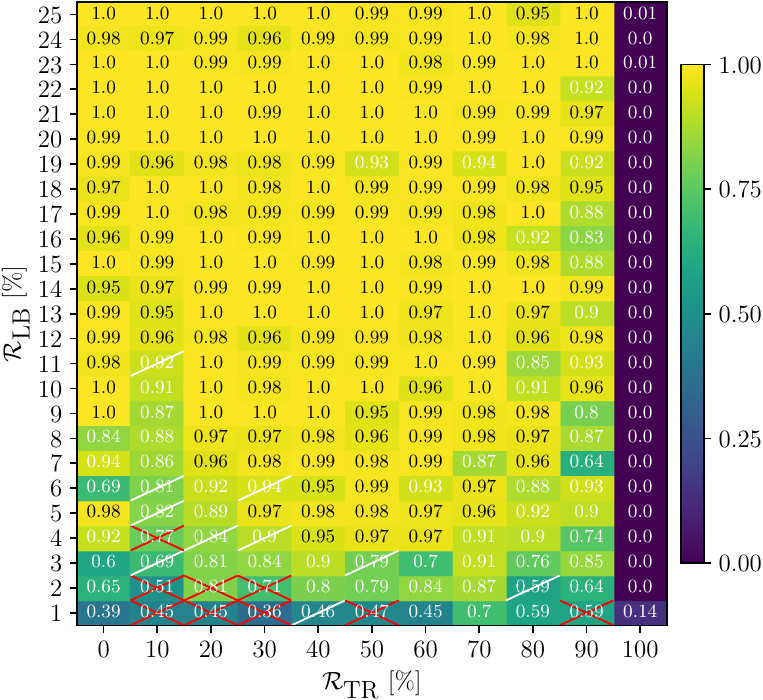}
    \label{fig:subfig:app-p2-cumulants-4}
  } 
  \caption{Supplementary Bhattacharyya coefficient $C_B$ maps for the
    $\mathcal{P}2$ kurtosis estimates.
    Panels (a) and (b) show the single-ensemble results for
    \texttt{L12T4b1.60k13585} obtained with the \ref{it:trace-feature}
    and \ref{it:plaquette-feature} approaches, respectively.
    Panel (c) shows an additional single-ensemble result for
    \texttt{L12T4b1.60k13582} obtained with the
    \ref{it:plaquette-feature} approach, and panel (d) shows the
    multi-ensemble reweighting result obtained with the same feature
    approach.
    The coefficient compares Gaussian summaries and is used
    descriptively; it is not a statistical-equivalence test or a
    criterion for establishing improved precision.}
  \label{fig:app-p2-cumulants}
\end{figure*}

Figure~\ref{fig:app-p2-cumulants} provides additional Bhattacharyya
coefficient $C_B$ maps to show how this behavior extends beyond the
representative case in Fig.~\ref{fig:app-p2-trace}.
The upper panels compare the \ref{it:trace-feature} and
\ref{it:plaquette-feature} approaches for the single-ensemble kurtosis
at $\kappa=0.13585$, while the lower panels show the single-ensemble
result at $\kappa=0.13582$ and the multi-ensemble reweighting result,
respectively.
These supplementary maps are included to document the observed
dataset- and analysis-dependent behavior of $\mathcal{P}2$, rather
than to define an additional estimator-selection criterion.

\begin{figure*}[tb]
  \centering
  \subfigure[\ref{it:plaquette-feature} approach]{
    \includegraphics[width=0.47\linewidth]{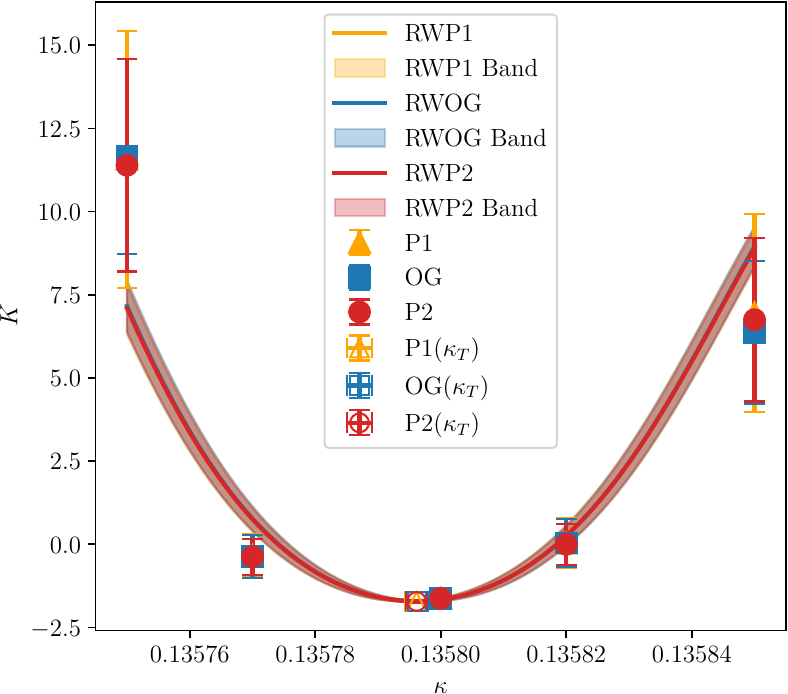}
    \label{fig:curve-ML2}
  }
  \hfill
  \subfigure[\ref{it:trace-feature} approach]{
    \includegraphics[width=0.47\linewidth]{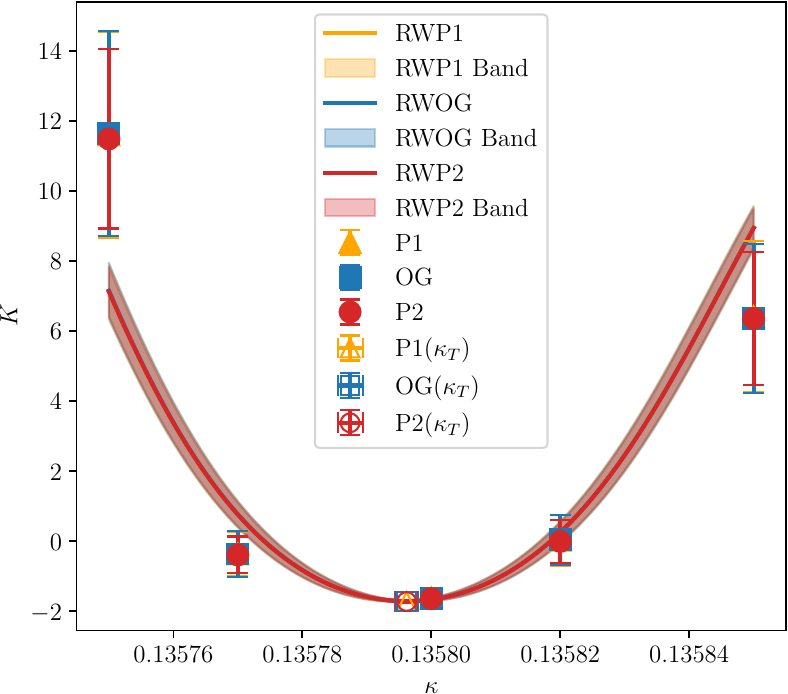}
    \label{fig:curve-ML1}
  }
  \caption{Representative multi-ensemble reweighting curves for the
    kurtosis.
    Blue, orange, and red denote the original, $\mathcal{P}1$, and
    $\mathcal{P}2$ results, respectively.
    Filled markers indicate single-ensemble evaluations, open markers
    show the interpolated transition points, and shaded bands show the
    associated uncertainties.
    Panel (a) uses $\mathcal{R}_{\mathrm{LB}}=15\,\%$ and
    $\mathcal{R}_{\mathrm{TR}}=20\,\%$; panel (b) uses
    $\mathcal{R}_{\mathrm{LB}}=1\,\%$ and
    $\mathcal{R}_{\mathrm{TR}}=10\,\%$.}
  \label{fig:app-p2-curves}
\end{figure*}

Fig.~\ref{fig:app-p2-curves} presents representative reweighting
curves retaining the direct visual comparison among the original,
$\mathcal{P}1$, and $\mathcal{P}2$ results.
The two panels illustrate different feature approaches and different
choices of the labeled and training fractions.
They are included to document the behavior of the alternative
estimator at the level of the reweighted curves and are not used to
support the primary cost projection.

\section{Detailed workflow of the multi-ensemble reweighting}
\label{sec:app-multi-rw}

The multi-ensemble reweighting procedure adopted in this study follows
the workflow summarized below, with individual steps denoted as
\textbf{R-}$\bm{x}$ ($x=1,\dots,12$).
\begin{enumerate}[
    label=\textbf{R-\arabic*}, ref=\textbf{R-\arabic*}, itemsep=0.5pt
  ]
\item \label{it:R-1} Load $\Tr \, M^{-j}$ ($j=1,2,3,4$) from ensembles
  that share the same lattice volume $V = N_{\text{S}}^3 \times
  N_{\text{T}}$ and inverse coupling constant $\beta$, but have
  different values of $\kappa$.
  Specifically, this refers to the subsets $S^Y_{\text{TR}}$,
  $S^Y_{\text{BC}}$ and $S^Y_{\text{UL}}$ that were divided in
  Sec.~\ref{sec:dataset-split}.
  In this step, we also load the ML-estimated counterparts
  $S^P_{\text{BC}}$ and $S^P_{\text{UL}}$.
\item \label{it:R-2} Reconstruct each loaded subset so that its
  configuration indices exactly match those in the original datasets
  (say, $S^Y$) of each ensemble.
  This process results in the formation of the following four
  $\mathcal{S}_i$:
  \begin{align}
    \mathcal{S}_1 &= S^Y_{\text{TR}} \cup S^Y_{\text{BC}} \cup
    S^Y_{\text{UL}} \quad\, \equiv \; S^Y \,, \label{eq:EA-1} \\
    \mathcal{S}_2 &= S^Y_{\text{TR}} \cup S^Y_{\text{BC}} \cup
    S^P_{\text{UL}} \,, \label{eq:EA-2} \\
    \mathcal{S}_3 &= S^Y_{\text{TR}} \cup S^Y_{\text{BC}} \qquad\qquad
    \equiv \; S^Y_{\text{LB}} \,, \label{eq:EA-3} \\
    \mathcal{S}_4 &= S^Y_{\text{TR}} \cup S^P_{\text{BC}}
    \,, \label{eq:EA-4}
  \end{align}
\item \label{it:R-3} For the computation of the reweighting factors,
  the free-energy offsets are evaluated using the Newton-Raphson
  method for all of the $\mathcal{S}_{i}$ ($i=1,2,3,4$) defined in
  Eqs.~\eqref{eq:EA-1}--\eqref{eq:EA-4}.
  For the detailed definition of the free-energy offset and its role
  in the multi-ensemble reweighting procedure, refer to
  Appendix~\ref{sec:free-energy-offset} and
  Appendix~\ref{sec:rw-factor}.
\item \label{it:R-4} Let $\kappa_I$ and $\kappa_F$ denote the initial
  and final hopping parameters, respectively, for the multi-ensemble
  $\kappa$-reweighting.
  Define the number of $\kappa$ points along the trajectory as
  $n_{\kappa}$.
  The interval between successive $\kappa$ values is given by
  \begin{equation}
    \delta \kappa = \frac{\kappa_F - \kappa_I}{n_{\kappa} - 1} \,,
  \end{equation}
  and each target hopping parameter is defined as
  \begin{equation}
    \kappa_r = \kappa_I + (r - 1) \, \delta \kappa \quad (1 \le r \le
    n_\kappa) \,,
    \label{eq:kappa-alpha-1}
  \end{equation}
  where $\kappa_r = \kappa_T$ is taken as the target point for the
  reweighting calculation performed at each step along the trajectory.
\item \label{it:R-5} If this step is visited for the first time, start
  with $r = 1$ (\textit{i.e.}, $\kappa_r = \kappa_I$).
  If it is revisited after returning from \ref{it:R-11}, update $r
  \!\to\! r + 1$ (and thus $\kappa_r$) according to
  Eq.~\eqref{eq:kappa-alpha-1}.
  Then proceed as follows.
  Compute $w_{a;n}(\kappa_r)$, the reweighting factor, for all
  $\mathcal{S}_{i}$ defined in Eqs.~\eqref{eq:EA-1}--\eqref{eq:EA-4},
  using the free-energy offsets previously obtained in \ref{it:R-3}.
  For the detailed procedure of the reweighting-factor computation,
  refer to Appendix~\ref{sec:rw-factor}.
\item \label{it:R-6} Using the $\left[\Tr \,
  M^{-j}\right]_{a;n}(\kappa_a)$, previously measured quantities at
  the original simulation points, estimate $\left[\Tr \,
    M^{-j}\right]_{a;n}(\kappa_r)$ approximately.
  For the detailed procedure of this estimation, refer to
  Appendix~\ref{sec:trace-shift}.
\item \label{it:R-7} Using the $\left[\Tr \,
  M^{-j}\right]_{a;n}(\kappa_r)$ obtained in \ref{it:R-6}, compute
  $\left[ Q_j \right]_{a;n}(\kappa_r)$ and evaluate $\left[ w \; Q_j
  \right]_{a;n}(\kappa_r)$ by multiplying each $\left[ Q_j
  \right]_{a;n}(\kappa_r)$ with the corresponding reweighting factor
  $w_{a;n}(\kappa_r)$ computed in \ref{it:R-5}, configuration by
  configuration.
\item \label{it:R-8} Partition $w_{a;n}(\kappa_r)$ and $\left[ w \;
  Q_j \right]_{a;n}(\kappa_r)$ into the original training
  ($S^{Y}_{\text{TR}}$), bias correction ($S^{Y}_{\text{BC}}$ and
  $S^P_{\text{BC}}$), and unlabeled sets ($S^{Y}_{\text{UL}}$ and
  $S^P_{\text{UL}}$), preserving the configuration ordering exactly as
  in the original data, for all $\mathcal{S}_{i}$ defined in
  Eqs.~\eqref{eq:EA-1}--\eqref{eq:EA-4}.
  Among these, the subsets used to construct the ML predictions
  $\mathcal{P}1$ and $\mathcal{P}2$ are chosen as follows:
  \begin{itemize}
  \item $\mathcal{S}_1$ represents the original dataset used to
    reproduce the conventional calculation without any ML involvement.
    Unlike the other $\mathcal{S}_i$, it is not partitioned but
    retained in its entirety, serving as the direct reference against
    which the ML estimations $\mathcal{P}1$ and $\mathcal{P}2$ are
    validated.
  \item The predicted unlabeled set $S_{\text{UL}}^{P}$ used for
    computing prediction~$\mathcal{P}1$ is taken from $\mathcal{S}_2$.
  \item The bias correction set $S_{\text{BC}}^{Y}$ used for computing
    prediction~$\mathcal{P}1$ is taken from $\mathcal{S}_3$.
  \item The predicted bias correction set $S_{\text{BC}}^{P}$ used for
    computing prediction~$\mathcal{P}1$ is taken from $\mathcal{S}_4$.
  \item The labeled set $S^Y_{\text{LB}} = S^Y_{\text{TR}} \cup
    S^Y_{\text{BC}}$ used for computing prediction~$\mathcal{P}2$ is
    taken from $\mathcal{S}_3$.
  \end{itemize}
\item \label{it:R-9} For each previously partitioned subset, perform
  bootstrap resampling.
  Within each subset, generate bootstrap replicas of $\left[w \,
    Q_j\right](\kappa_r)$ and $w(\kappa_r)$ using identical resampling
  indices, and then compute the reweighted observables by dividing
  each pair of corresponding replicas of $\left[w \,
    Q_j\right](\kappa_r)$ by those of $w(\kappa_r)$.
  As a result, the bootstrap replicas of the reweighted ensemble
  averages at the target $\kappa_r$ are obtained for all subsets.
  Then, combine these replicas across subsets and evaluate $\langle
  Q_j \rangle_{\mathcal{P}1}$ [Eq.~\eqref{eq:P1-1}] and $\langle Q_j
  \rangle_{\mathcal{P}2}$ [Eq.~\eqref{eq:P2-1}].
\item \label{it:R-10} Using $\langle Q_j \rangle_{\mathcal{P}1}$
  obtained in \ref{it:R-9}, compute the cumulants of the chiral
  condensate --- from the condensate itself to the kurtosis ---
  according to Eqs.~\eqref{eq:cond}--\eqref{eq:kurt}, performing the
  calculation pairwise over bootstrap replicas.
  The same procedure is applied to $\mathcal{P}2$.
\item \label{it:R-11} Evaluate the mean values and statistical
  uncertainties using the bootstrap replicas obtained in the previous
  steps.
  Afterward, if the current target satisfies $\kappa_r < \kappa_F$,
  return to \ref{it:R-5}.
  If $\kappa_r = \kappa_F$, move on to \ref{it:R-12} for the
  subsequent interpolation step.
\item \label{it:R-12} Having completed the multi-ensemble reweighting
  calculations for all $\kappa_I \le \kappa_r \le \kappa_F$, we
  construct the reweighting-curve bands.
  At this stage, we perform a three-point Lagrange interpolation along
  the $\kappa$ trajectory, taking each transition-point candidate
  identified from the bootstrap replicas as the central point and
  including its two neighboring points on both sides.
  Finally, statistical uncertainties of the interpolated results are
  obtained from the distribution of bootstrap replicas.
\end{enumerate}

At this point, one might also wonder whether, following an alternative
construction of \ref{it:R-2} in which $\mathcal{S}_5$ is defined as
\begin{align}
  \mathcal{S}_5 &= S^Y_{\text{TR}} \cup S^P_{\text{BC}} \cup
  S^P_{\text{UL}}\,, \label{eq:EA-5}
\end{align}
the representative subsets could simply be chosen as follows:
\begin{itemize}
\item $S^P_{\text{UL}}$ for prediction $\mathcal{P}1$ taken from
  $\mathcal{S}_2$;
\item $S^{Y}_{\text{BC}}$ for prediction $\mathcal{P}1$ also taken
  from $\mathcal{S}_2$;
\item $S^P_{\text{BC}}$ for prediction $\mathcal{P}1$ taken from
  $\mathcal{S}_5$;
\item $S^Y_{\text{LB}} = S^Y_{\text{TR}} \cup S^Y_{\text{BC}}$ for
  prediction $\mathcal{P}2$ taken from $\mathcal{S}_2$.
\end{itemize}
However, if one proceeds in this way, then—after the computation of
the free-energy offsets in \ref{it:R-3} and of the reweighting factors
in \ref{it:R-5}, followed by the subset repartitioning in
\ref{it:R-8}—the partitions of $w_{a;n}(\kappa_r)$ and $[\,w \,
  Q_j\,]_{a;n}(\kappa_r)$ are, by construction, subject to
\emph{prediction-induced inaccuracy} carried over from
$S^P_{\text{UL}}$.
Accordingly, the quantities corresponding to $S^{Y}_{\text{BC}}$,
$S^P_{\text{BC}}$, and $S^{Y}_{\text{LB}}$ may be affected by
additional inaccuracies propagated from $S^P_{\text{UL}}$; this
propagation is structural to the construction.
As a consequence, immediately after the reweighting, the computation
of prediction~$\mathcal{P}1$ may fail to realize a bias correction
that is as faithful to its intended role as originally designed.
Moreover, even the computation of prediction~$\mathcal{P}2$—whose
stated goal is to improve statistical precision via a weighted average
with the labeled set—faces the risk of contamination by these
carried-over inaccuracies from $S^P_{\text{UL}}$.
Accordingly, in the present work we construct index-aligned datasets
in \ref{it:R-2}---at full cardinality for $\mathcal{S}_1$ and
$\mathcal{S}_2$, and as index-aligned, labeled-sector subsets for
$\mathcal{S}_3$ and $\mathcal{S}_4$---and select the representative
subsets as prescribed in \ref{it:R-8}.

As shown in Eqs.~\eqref{eq:cond}--\eqref{eq:C_4}, the cumulants are
constructed from the ensemble averages of $Q_i$, namely $\langle Q_i
\rangle$, which represent the moments.
This is the reason why, in steps~\ref{it:R-9} and \ref{it:R-10}, the
reweighted observables are evaluated at the level of $Q_i$.
Accordingly, the reweighted ensemble averages at a given target
hopping parameter $\kappa_T$, defined in Eq.~\eqref{eq:rw-define-1},
are evaluated based on these $Q_i$ values.

\section{Computation of free-energy offsets for multi-ensemble reweighting}
\label{sec:free-energy-offset}

To perform such a multi-ensemble reweighting analysis, we must compute
the reweighting factors required to estimate the moments (quark-loop
contributions) and the resulting cumulants at intermediate $\kappa$
values where direct measurements are unavailable.
For this purpose, the free-energy offsets $\{ f_b\}$ for each ensemble
need to be determined in advance by solving the MBAR (Multistate
Bennett Acceptance Ratio) self-consistency
equations~\cite{Bennet:1976aa, Shirts:2008jcp, Kuramashi:2016kpb}.

The MBAR method provides a statistically optimal estimator—minimum
variance and asymptotically unbiased—for the relative free energies
among multiple ensembles, assuming finite sampling from each.
In this work, the MBAR equations are solved using a Newton-Raphson
iterative solver.

We consider $R$ ensembles indexed by $a = 1, \dots, R$, each
containing $N_a$ configurations and characterized by a simulation
parameter (\textit{e.g.}, $\beta_a$ or $\kappa_a$) denoted $\theta_a$.
However, since the present study focuses exclusively on reweighting
with respect to the hopping parameter $\kappa$, we hereafter write
$\kappa_a$ explicitly instead of the generic notation $\theta_a$.
For each gauge configuration $\mathcal{U}_{a;n}$ drawn from the $a$-th
ensemble (where $n$ labels the configuration index for certain $a$-th
ensemble), the action evaluated at parameter $\kappa_a$ is denoted
$S(\kappa_a;\mathcal{U}_{a;n})$, and we define the action shift
\begin{equation}
  \Delta S_{a;n}(\kappa_b) \equiv S(\kappa_b;\mathcal{U}_{a;n}) -
  S(\kappa_a;\mathcal{U}_{a;n})
\end{equation}
where its explicit aspect is discussed in
Appendix~\ref{sec:quark-mass-action-shift}.
Fixing one ensemble (typically the last, $a=R$) as the reference, we
solve for the free-energy offset vector
\begin{equation}
  \bm{f} = (f_1, f_2, \dots, f_{R-1})^{\mathsf{T}} \, .
\end{equation}

Before performing the Newton-Raphson iteration, it is necessary to
provide a reasonable initial guess for the free energy offsets
$\bm{f}$ to ensure numerical stability.
For each ensemble $b$, the initial guess $\{ f_b^{(0)} \}$ are set to
the maximum action shift $\Delta S_{a;n}(\kappa_b)$ evaluated over all
configurations $n$ of all ensembles $a$, under the parameter
$\kappa_b$:
\begin{equation}
  f_b^{(0)} = \max_{1 \le a \le R} \left( \max_{1 \le n \le N_a}
  \Delta S_{a;n}(\kappa_b) \right) \,.
\end{equation}
Choosing the maximum value provides a safe initialization that
prevents early underflow of exponential weights, although the
iteration quickly relaxes toward the self-consistent region.

For each target ensemble $b = 1, \dots, R - 1$, we define
\begin{align}
  D_{a;n}^{(b)}(f) &= \sum_{d=1}^{R} N_d \; \exp \left[
    E_{a;n}^{(b,d)}(f) \right] \,,
\end{align}
where
\begin{align}
  E_{a;n}^{(b,d)}(f) &= \Delta S_{a;n}(\kappa_b) - \Delta
  S_{a;n}(\kappa_d) - f_b + f_d \,.
\end{align}
The normalization sum for each $b$ is then given by
\begin{equation}
  T_b(f) = \sum_{a=1}^{R} \sum_{n=1}^{N_a} \frac{1}{D_{a;n}^{(b)}(f)}
  \, .
\end{equation}
The MBAR residual equations take the form
\begin{equation}
  F_b(f) = \log T_b(f) = 0, \qquad b = 1 \,, \dots\,,\; R - 1 \,,
  \label{eq:resid-F-1}
\end{equation}
which enforces the self-consistency condition $T_b(f) = 1$,
corresponding to the statistical equilibrium among all ensembles in
the reweighting framework.

For the Newton-Raphson iteration, we compute the Jacobian entries
\begin{align}
  J_{bc}(f) & = \frac{\partial F_b(f)}{\partial f_c} \nonumber \\
  & = \frac{1}{T_b(f)} \; \sum_{a=1}^{R} \sum_{n=1}^{N_a} \left(
  -\frac{1}{\left[ D_{a;n}^{(b)}(f) \right]^2} \frac{\partial
    D_{a;n}^{(b)}}{\partial f_c} \right) \,.
  \label{eq:J-1}
\end{align}
Here, the derivative of $D_{a;n}^{(b)}(f)$ with respect to $f_c$ is
\begin{equation}
  \frac{\partial D_{a;n}^{(b)}}{\partial f_c} = N_c \; \exp\left[
    E_{a;n}^{(b,c)}(f) \right] - \delta_{bc} \; D_{a;n}^{(b)}(f)
  \,. \label{eq:dDdf-1}
\end{equation}
Substituting Eq.~\eqref{eq:dDdf-1} into Eq.~\eqref{eq:J-1} gives
\begin{equation}
  J_{bc}(f) = \frac{-1}{T_b(f)} \sum_{a=1}^{R} \sum_{n=1}^{N_a}
  \frac{N_c \; \exp\left[E_{a;n}^{(b,c)}(f)\right]}{\left[
      D_{a;n}^{(b)}(f) \right]^2} + \delta_{bc} \, .
    \label{eq:Jacob-J-1}
\end{equation}

Once the residual vector $\mathbf{F}(f)$ (Eq.~\eqref{eq:resid-F-1})
and its Jacobian matrix $J(f)$ (Eq.~\eqref{eq:Jacob-J-1}) have been
constructed, the nonlinear system is solved iteratively using
Newton–Raphson updates:
\begin{equation}
  \bm{f} \;\mapsto\; \bm{f} - J^{-1}(f) \; \mathbf{F}(f) \, .
\end{equation}
This iterative process continues until the residuals $F_b(f)$ fall
below a specified tolerance, yielding the self-consistent free-energy
offsets $f_b$ to be used for computing the ensemble reweighting
factors.

\section{Computation of reweighting factors}
\label{sec:rw-factor}

Once the self-consistent free-energy offsets $\{ f_b \}$ are obtained,
the normalized reweighting factors $w_{a;n}$ for each configuration
$\mathcal{U}_{a;n}$ can be computed for an arbitrary target hopping
parameter $\kappa_T$.\footnote{Here, $\kappa_T$ should not be confused
with $\kappa_t$. The symbol $\kappa_T$ indicates the ``target''
hopping parameter used in the multi-ensemble reweighting procedure,
whereas $\kappa_t$ denotes the hopping parameter at the transition
point on the phase structure.}
Each configuration belonging to $a$-th ensemble with configuration
index $n$ is assigned a weight that incorporates both the action
shifts $\Delta S_{a;n}(\kappa_T)$, $\Delta S_{a;n}(\kappa_b)$, and the
previously determined free-energy offsets $f_b$.
To avoid numerical overflow or underflow during the evaluation of
exponentials, the computation is stabilized using the
\textit{log-sum-exp} trick implemented through $\mathcal{X}$.
The weights are thus defined as
\begin{equation}
  w_{a;n}(\kappa_T) = \frac{\exp \!\left[ - \Delta S_{a;n}(\kappa_T)
      \right]}{ \displaystyle\sum_{b=1}^{R} N_b \; \exp \!\left[ -
      \Delta S_{a;n}(\kappa_b) + f_b - \mathcal{X} \right] } \,,
\end{equation}
where $\mathcal{X}$ is defined as
\begin{align}
  \mathcal{X} &= \max_{1 \le a,b \le R} \left( \max_{1 \le n \le N_a}
  \mathcal{X}_{a,b;n} \right) \,, \\
  \mathcal{X}_{a,b;n} & = \Delta S_{a;n}(\kappa_T) - \Delta
  S_{a;n}(\kappa_b) + f_b \,.
\end{align}
Subtracting $\mathcal{X}$ from all exponents leaves
$w_{a;n}(\kappa_T)$ unchanged, because the constant cancels between
the numerator and denominator, but it shifts all intermediate
exponentials into a numerically safe range.

It should also be noted that $\exp \left[ -\,\Delta S_{a;n}(\kappa_T)
  \right]$ originates from the quark determinant:\footnote{Note that
we are considering $\kappa$-reweighting itself, for which the gluonic
part of the action does not contribute at all and cancels
automatically between the numerator and denominator.}
\begin{equation}
  \exp \left[ -\,\Delta S_{a;n}(\kappa_T) \right] = \left[ \left(
    \frac{\det M(\kappa_T)}{\det M(\kappa_a)} \right)^{N_{\text{f}}}
    \right]_{a;n} \,,
\end{equation}
which follows directly from the quark part of the action (see also
Eq.~\eqref{eq:det-ratio-1}).
A detailed derivation is provided in
Appendix~\ref{sec:quark-mass-action-shift}.

With these stabilized weights, the expectation value of an observable
$\Omega$ at the target parameter $\kappa_T$ is estimated as
\begin{equation}
  \left\langle \Omega(\kappa_T;\mathcal{U}) \right\rangle =
  \frac{\displaystyle \sum_{a=1}^{R} \sum_{n=1}^{N_a}
    w_{a;n}(\kappa_T) \; \Omega(\kappa_T;\mathcal{U}_{a;n})}
       {\displaystyle \sum_{a=1}^{R} \sum_{n=1}^{N_a}
         w_{a;n}(\kappa_T)} \, .
  \label{eq:rw-define-1}
\end{equation}
This reweighting procedure enables the evaluation of observables at
parameter values where no direct measurements were performed, by
optimally combining configurations from multiple ensembles through the
self-consistently determined free-energy offsets $\{ f_b \}$.

It is worth noting that, in Eq.~\eqref{eq:rw-define-1}, the
observables actually measured in simulations are
$\Omega(\kappa_a;\mathcal{U}_{a;n})$, whereas no direct data exist for
$\Omega(\kappa_T;\mathcal{U}_{a;n})$ at the target hopping parameter
$\kappa_T$.  In practice, we approximate the latter using a Taylor
expansion of $\Omega(\kappa_a;\mathcal{U}_{a;n})$, as discussed in
Appendix~\ref{sec:trace-shift}.

For compactness of notation, we hereafter denote
$\Omega_{a;n}(\kappa_T) \equiv \Omega(\kappa_T;\mathcal{U}_{a;n})$.

\section{Trace shift via Taylor expansion}
\label{sec:trace-shift}

To evaluate $\Tr \, M^{-k}$ at a target hopping parameter $\kappa_T$
close to a simulation point $\kappa_a$, we employ a Taylor expansion
of the $\Tr \, M^{-k}$ with respect to the quark mass.
This provides a controlled fourth-order approximation to $\Tr \,
M^{-k}$ at $\kappa_T$, allowing smooth interpolation of quark-loop
observables without performing additional Dirac-operator inversions.

In what follows, we present a quark-mass-based formulation that is
convenient for the expansion.
Up to Eq.~\eqref{eq:general-shift}, we temporarily use the quark mass
$m$
\begin{equation}
  a\,m = \frac{1}{2\kappa} + \text{const.}\,,
\end{equation}
instead of the hopping parameter $\kappa$, setting the lattice spacing
to $a=1$ for notational simplicity.
Thereafter, we return to the $\kappa$ representation.

With the mass shift\footnote{Although the mass shift is conventionally
defined as $m_T - m_a$, we intentionally adopt the opposite sign,
$\delta m_{T,a} = m_a - m_T$, to simplify later expressions by
avoiding redundant negative signs that would otherwise appear
throughout the derivation.}
\begin{equation}
  \delta m_{T,a} = m_a - m_T = \frac{1}{2} \left( \frac{1}{\kappa_a} -
  \frac{1}{\kappa_T} \right) \,,
    \label{eq:delta-m-1}
\end{equation}
where the hopping parameter $\kappa$ is explicitly shown for clarity,
the Taylor expansion for an observable $\Omega$ around $m_a$ reads
\begin{equation}
  \Omega_{a;n}(m_T) = \sum_{j=0}^{\infty} \frac{(-\delta
    m_{T,a})^{j}}{j!}  \left(\frac{\partial}{\partial m_a}\right)^{j}
  \Omega_{a;n}(m_a) \,.
  \label{eq:taylor-omega}
\end{equation}

To evaluate the series in Eq.~\eqref{eq:taylor-omega} for the specific
choice $\Omega=\Tr\,M^{-k}$, we require the following general identity
for repeated mass derivatives, which can be established by
mathematical induction:
\begin{align}
  \frac{\partial^{j}}{\partial m^{j}}\,\Tr\,M^{-k} &= (-1)^{j} \,
  \frac{\left(k+j-1\right)!}{\left(k-1\right)!} \,\Tr\,M^{-(k+j)} \,.
  \label{eq:deriv-rule}
\end{align}
We prove Eq.~\eqref{eq:deriv-rule} by induction on $j \ge 1$.
The case $j=1$ is shown by the proof of
Eq.~\eqref{eq:deriv-rule-base}.
Assume Eq.~\eqref{eq:deriv-rule} holds for $j$; then
\begin{align}
  & \hphantom{=} \frac{\partial^{j+1}}{\partial m^{j+1}}\,\Tr\,M^{-k}
  \nonumber \\
  &= \frac{\partial}{\partial m}\left[\left(-1\right)^{j} \,
    \frac{\left(k+j-1\right)!}{\left(k-1\right)!} \, \Tr \,
    M^{-\left(k+j\right)}\right] \nonumber\\
  &= \left(-1\right)^{j} \,
  \frac{\left(k+j-1\right)!}{\left(k-1\right)!} \,
  \left(-\left(k+j\right) \, \Tr \,
  M^{-\left(k+j+1\right)}\right)\nonumber\\
  &= \left(-1\right)^{j+1} \,
  \frac{\left(k+j\right)!}{\left(k-1\right)!} \, \Tr \,
  M^{-\left(k+j+1\right)}\,,
\end{align}
which completes the induction.
This completes the proof of Eq.~\eqref{eq:deriv-rule}.

Substituting \eqref{eq:deriv-rule} into \eqref{eq:taylor-omega} with
$\Omega = \Tr \, M^{-k}$ gives the general mass-shift formula
\begin{align}
  \left[ \Tr\,M^{-k} \right]_{a;n}(m_T) &= \sum_{j=0}^{\infty}
  \frac{\left(k+j-1\right)!}{j! \; \left(k-1\right)!} \; (\delta
  m_{T,a})^{j} \nonumber \\
  &\hphantom{=} \quad \times \left[\Tr\,M^{-(k+j)}\right]_{a;n}(m_a)
  \,.
  \label{eq:general-shift}
\end{align}
Since $(-\delta m_{T,a})^{j} \, (-1)^{j} = (\delta m_{T,a})^{j}$,
Eq.~\eqref{eq:general-shift} follows directly from
\eqref{eq:taylor-omega}.
A sufficient condition for convergence is $|\delta
m_{T,a}|\,\|M(\kappa_a)^{-1}\|<1$ (\textit{e.g.}, in any consistent
matrix norm).
Retaining terms up to the indicated orders, we obtain:
\begin{align}
  \left[\Tr \, M^{-1}\right]_{a;n}(\kappa_T) &= \left[\Tr \,
    M^{-1}\right]_{a;n}(\kappa_a) \nonumber \\
  &\hphantom{=} + \left( \delta m_{T,a} \right)^1 \; \left[\Tr \,
    M^{-2}\right]_{a;n}(\kappa_a) \nonumber \\
  &\hphantom{=} + \left( \delta m_{T,a} \right)^2 \; \left[\Tr \,
    M^{-3}\right]_{a;n}(\kappa_a) \nonumber \\
  &\hphantom{=} + \left( \delta m_{T,a} \right)^3 \; \left[\Tr \,
    M^{-4}\right]_{a;n}(\kappa_a) \nonumber \\
  &\hphantom{=} + \mathcal{O}\left(\left( \delta m_{T,a}
  \right)^4\right) \,, \\[0.3em]
  \left[\Tr \, M^{-2}\right]_{a;n}(\kappa_T) &= \left[\Tr \,
    M^{-2}\right]_{a;n}(\kappa_a) \nonumber \\
  &\hphantom{=} + 2 \; \left( \delta m_{T,a} \right)^1 \; \left[\Tr \,
    M^{-3}\right]_{a;n}(\kappa_a) \nonumber \\
  &\hphantom{=} + 3 \; \left( \delta m_{T,a} \right)^2 \; \left[\Tr \,
    M^{-4}\right]_{a;n}(\kappa_a) \nonumber \\
  &\hphantom{=} + \mathcal{O}\left(\left( \delta m_{T,a}
  \right)^3\right) \,, \\[0.3em]
  \left[\Tr \, M^{-3}\right]_{a;n}(\kappa_T) &= \left[\Tr \,
    M^{-3}\right]_{a;n}(\kappa_a) \nonumber \\
  &\hphantom{=} + 3 \; \left( \delta m_{T,a} \right) \; \left[\Tr \,
    M^{-4}\right]_{a;n}(\kappa_a) \nonumber \\
  &\hphantom{=} + \mathcal{O}\left(\left( \delta m_{T,a}
  \right)^2\right) \,, \\[0.3em]
  \left[\Tr \, M^{-4}\right]_{a;n}(\kappa_T) &= \left[\Tr \,
    M^{-4}\right]_{a;n}(\kappa_a) \nonumber \\
  &\hphantom{=} + \mathcal{O}\left(\left( \delta m_{T,a}
  \right)^1\right) \,.
\end{align}

\section{Action shift via Taylor expansion}
\label{sec:quark-mass-action-shift}

The action shift
\begin{equation}
  \Delta S_{a;n}(\kappa_T) \equiv S(\kappa_T;\mathcal{U}_{a;n}) -
  S(\kappa_a;\mathcal{U}_{a;n})
\end{equation}
associated with a change in the hopping parameter $\kappa$ can be
estimated by a Taylor expansion in the mass variable around the
original simulation point.
This expansion enables the evaluation of reweighting factors for
nearby target values $\kappa_T$, without generating new ensembles at
$\kappa_T$.

We start from the identity
\begin{equation}
  \left[ \left(\frac{\det M(\kappa_T)}{\det
      M(\kappa_a)}\right)^{N_{\text{f}}} \right]_{a;n} = \exp \left[
    N_{\text{f}}\, L_{a;n} \right] \,,
\end{equation}
where
\begin{equation}
  L_{a;n} = \left[ \Tr \, \{\ln M(\kappa_T) - \ln M(\kappa_a)\}
    \right]_{a;n} \,.
\end{equation}
The Dirac operator depends linearly on the quark mass, so a small
change in $\kappa$ induces a scalar shift of the mass term:
\begin{equation}
  M(\kappa_T) = M(\kappa_a) - \delta m_{T,a}\,\mathbb{1}
\end{equation}
where $\delta m_{T,a}$ is defined in Eq.~\eqref{eq:delta-m-1}.
Using the factorization
\begin{equation}
  M(\kappa_T) = M(\kappa_a) \left[ \mathbb{1} - \delta m_{T,a} \;
    M(\kappa_a)^{-1}\right]
\end{equation}
and the fact that the two factors commute, we obtain
\begin{align}
  L_{a;n} &= \left[ \Tr \, \ln \left( \mathbb{1} - \delta
    m_{T,a}\,M(\kappa_a)^{-1} \right) \right]_{a;n}.
\end{align}
For sufficiently small $\left\lvert\delta m_{T,a}\right\rvert$ (so
that the series converges), the matrix logarithm admits the power
series
\begin{equation}
  \ln(\mathbb{1}-X) = -\sum_{j=1}^{\infty}\frac{X^{j}}{j}, \quad X =
  \delta m_{T,a}\,M(\kappa_a)^{-1} \,.
  \label{eq:ln-1-x-1}
\end{equation}
Taking the trace to Eq.~\eqref{eq:ln-1-x-1} and substituting the
expansion gives
\begin{equation}
  L_{a;n} = -\sum_{j=1}^{\infty}\frac{(\delta m_{T,a})^{j}}{j}\,
  \left[\Tr\,M^{-j}\right]_{a;n}(\kappa_a) \,.
\end{equation}
Combining the above, the determinant ratio can be written as
\begin{equation}
  \left[ \left(\frac{\det M(\kappa_T)}{\det
      M(\kappa_a)}\right)^{N_{\text{f}}} \right]_{a;n} = \exp\!\left[
    -\,\Delta S_{a;n}(\kappa_T) \right] \,,
    \label{eq:det-ratio-1}
\end{equation}
with the quark action shift
\begin{align}
  \Delta S_{a;n}(\kappa_T) &= N_{\text{f}} \; \sum_{j=1}^{4}
  \frac{(\delta m_{T,a})^{j}}{j}\,
  \left[\Tr\,M^{-j}\right]_{a;n}(\kappa_a) \nonumber \\
  &\hphantom{=} + \mathcal{O}\left( \left( \delta m_{T,a} \right)^{5}
  \right) \,.
\end{align}

\section{Block Bootstrap resampling}
\label{sec:block-bs}

As discussed in Sec.~\ref{sec:stat-err}, the data analyzed in this
study are taken in the vicinity of a first-order phase transition,
where autocorrelations are known to be strong.
One way to mitigate the potential underestimation of statistical
uncertainties caused by such dependence is to employ the
non-overlapping block bootstrap method~\cite{Carlstein:1986zz}.

In this procedure, we vary the block size from $1$ (corresponding to
the \textit{i.i.d.}~bootstrap) up to $1000$ along the horizontal axis,
while the vertical axis represents representative observables such as
the plaquette, rectangle, or $\Tr\,M^{-1}$.
These observables are mainly used as input features under the
assumption that their correlations with the target quantities are
sufficiently strong.

Given that the present dataset consists of $N = 20{,}000$ samples (see
Table~\ref{tab:ensemble-1}), we set the maximum block size to
$B_{\max} \approx N / 20 = 1000$, ensuring that at least about twenty
blocks are available for resampling.
For each block size $B$, we compute the noise-to-signal ratio
$R^{\text{NS}}_B$ at the block size $b=B$
\begin{equation}
  R^{\text{NS}}_B = \left. \frac{\sigma_X}{\bar{X}} \right|_{b = B}
  \,,
\end{equation}
and examine how $R^{\text{NS}}_B$ varies with the block size $B$ to
identify the region where it becomes approximately stable.
However, the resulting dependence on the block size $B$ tends to
exhibit considerable irregular oscillations rather than a smooth
curve, making it difficult to identify an optimal block size directly.

\begin{figure}[tb]
  \includegraphics[width=0.97\linewidth]{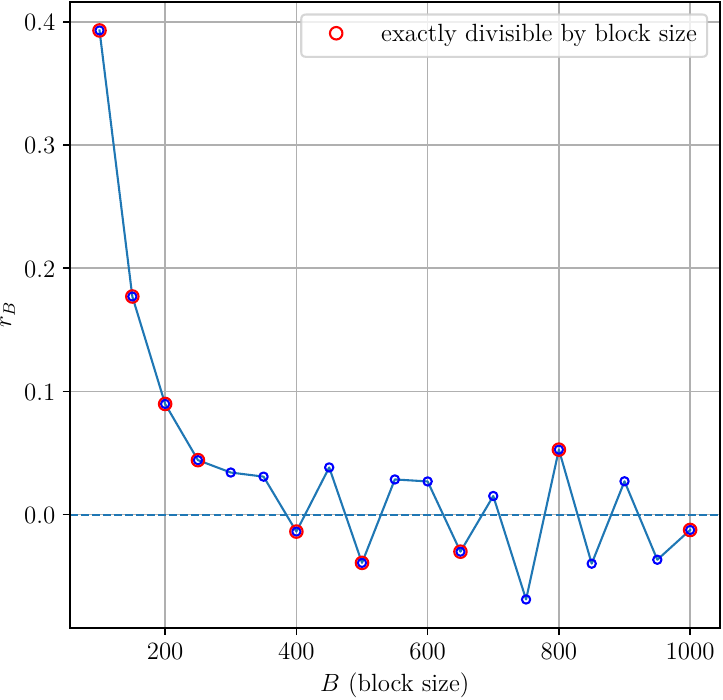}
  \caption{ Residual-like quantity $r_B$ [Eq.~(\ref{eq:block-resid})]
    for the plaquette as a function of block size $B$ used for
    determining an appropriate block length in the non-overlapping
    block bootstrap analysis.
    A moving-average smoothing with a window size of $50$ is applied
    to reduce short-scale fluctuations.
    The red markers indicate block sizes for which the total number of
    samples ($N = 20{,}000$) is exactly divisible by $B$, allowing
    full data utilization without any discarded remainder.
    Around $B \simeq 400$, $r_B$ occasionally drops below zero,
    suggesting that the effective noise level becomes nearly
    stationary, which we take as an empirical indicator for a
    reasonable block size.}
  \label{fig:block-scan}
\end{figure}  

To address this issue, we introduce a coarse-graining procedure: a
moving average is computed over a fixed window (chosen to be $50$ in
this study) to smooth out short-scale fluctuations.
We then compute a residual-like quantity, defined as the normalized
difference between successive points of the noise-to-signal ratio at
block size $b=B$,
\begin{equation}
  r_B = \frac{R^{\text{NS}}_{B} - R^{\text{NS}}_{B -
      1}}{R^{\text{NS}}_{B}} \,, \label{eq:block-resid}
\end{equation}
to quantify local changes.
The plot presented in Fig.~\ref{fig:block-scan} shows these residuals
as a function of block size.

The red markers in the figure indicate block sizes for which the total
number of samples is exactly divisible by the block size, meaning that
all data are used without any discarded remainder in the
non-overlapping block bootstrap procedure.
As such, we preferably choose block sizes near these points to
maximize data utilization.

From the plot, one can see that around a block size $B$ of $400$, the
residual quantity occasionally drops below zero, suggesting a possible
stabilization of the effective noise level.
We use this sign change only as an empirical selection heuristic; it
does not establish stationarity or identify an optimal block length.

Although a more rigorous determination of the optimal block size would
require further dedicated studies, in this work we employ the above
empirical procedure to estimate a reasonable block length which we
then use as an input parameter for our analysis.
In general, the determination of an optimal block size remains a
difficult problem, and no universally rigorous method has yet been
established.

Our present approach therefore provides only a rough empirical
estimate of the block size, which we use for practical purposes in
this analysis; the reported uncertainties are conditional on this
choice.
A more systematic and rigorous determination would require further
validation and extensive numerical studies in the future.

\section{About Deborah.jl package}
\label{sec:deborah}

To verify the accuracy of the bias-corrected ML estimation presented
in this work, we have developed a Julia package, \texttt{Deborah.jl}
\cite{Choi:2026zen}.
The name ``\texttt{Deborah.jl}'' stands for ``\textit{Deborah.jl is an
  Estimation tool for Bias-cOrrected Regression Analysis with
  Heuristics}'', which is a recursive acronym.
This appendix provides a brief overview of its main modules and core
functionalities for clarity and practical application.

\subsection{DeborahCore module}

The \texttt{DeborahCore} module corresponds to the computation
described in Sec.~\ref{sec:trace}.
Originally, this functionality existed as an independent package
dedicated solely to bias-corrected ML estimation.
However, to provide a unified framework encompassing not only the ML
estimation of the traces of the inverse Dirac operator but also
additional functionalities— such as cumulant estimation for single
ensembles [Sec.~\ref{sec:cumulant-single}] and multi-ensemble
reweighting [Sec.~\ref{sec:cumulant-reweighting}]—all related
components were integrated into a single package, with the original
core functionality retained under the submodule name
\texttt{DeborahCore}.
Although the present work specifically employs \texttt{DeborahCore}
for the bias-corrected ML estimation of $\Tr \, M^{-n}$, the module is
designed to be largely model- and observable-agnostic, allowing
flexible use regardless of the specific choice of features or targets.
As illustrated in the \ref{it:plaquette-feature} approach, where both
the plaquette and rectangle are used as features, the module supports
multiple input features.
While the calculations presented in Sec.~\ref{sec:trace} employed
\texttt{LightGBM} \cite{Ke:2017zz} as the default ML model,
\texttt{DeborahCore} also supports alternative models such as
\textsc{Lasso} \cite{Santosa:1986zz, Tibshirani:2018zz} and
\textsc{Ridge} \cite{Hoerl:1970aa, Hoerl:1970bb} regression.

\subsection{Esther module}

The \texttt{Esther} module corresponds to the computation described in
Sec.~\ref{sec:cumulant-single}.
It operates with the assistance of an auxiliary module,
\texttt{DeborahEsther}, which performs a preliminary check to ensure
that all required traces $\Tr \, M^{-n}$ have completed ML
estimations.
If any are missing, the module requests \texttt{DeborahCore} to
perform the necessary estimations first; once all data become
available, it proceeds to compute the cumulants for the single
ensemble.
The name ``\texttt{Esther}'' stands for ``\textit{Esther is a Summary
  Tool for Higher-order cumulants through Estimation via
  Regression}'', which is also a recursive acronym.

\subsection{Miriam module}

The \texttt{Miriam} module corresponds to the computation described in
Sec.~\ref{sec:cumulant-reweighting}.
It operates in conjunction with an auxiliary module,
\texttt{DeborahEstherMiriam}, which performs a preliminary check
across ensembles at different hopping parameters~$\kappa$ to ensure
that all required ML estimations of $\Tr \, M^{-n}$ have been
completed.
If any of them are missing, the module delegates the task to
\texttt{DeborahEsther}, which acts as an intermediary manager that
coordinates \texttt{DeborahCore} to perform the required ML
regressions for each corresponding ensemble.
Once all trace data become available, \texttt{Miriam} collects them
from the various ensembles and performs ML cumulant estimation through
multi-ensemble reweighting.
The name ``\texttt{Miriam}'' stands for ``\textit{MultI-ensemble
  Reweighting and Interpolation Analysis with Miriam}'', again a
recursive acronym.

\subsection{Sarah module}

The \texttt{Sarah} module provides a collection of shared utilities
and abstractions that are commonly used by the other core modules,
including \texttt{DeborahCore}, \texttt{Esther}, and \texttt{Miriam}.
It serves as a centralized hub for reusable auxiliary functions and
data structures, ensuring consistency and reducing code duplication
across the package.

\subsection{Rebekah module}

The \texttt{Rebekah} module collectively handles the generation of
figures and diagnostic plots that summarize and visualize the results
obtained from the main computational modules, including
\texttt{DeborahCore}, \texttt{Esther}, and \texttt{Miriam}.
It serves as a comprehensive reporting and evaluation layer that
aggregates outputs across different stages of the workflow, providing
benchmarking and explainable visual summaries of the bias-corrected ML
estimations and the resulting cumulants.
Figs.~\ref{fig:trace-heatmap-ML1-LBP-1-25}--\ref{fig:curve-ML1}, which
display the ML estimation results discussed in this paper, were
generated using the \texttt{Rebekah} module.

\subsection{Elijah module}

The \texttt{Elijah} module provides an interactive interface for
generating configuration files required by the main computational
modules, including \texttt{DeborahCore}, \texttt{Esther}, and
\texttt{Miriam}.
While these modules accept input parameters via \texttt{.toml}
configuration files, \texttt{Elijah} allows users to specify all
necessary parameters through a question--answer--style dialogue,
automatically creating valid \texttt{.toml} files that can be directly
used to launch the corresponding computations.
In this sense, it functions as a typical \textit{wizard} that assists
users in setting up and initiating the desired calculations.

\subsection{Rahab module}

The \texttt{Rahab} module provides diagnostic and preprocessing
utilities that are used prior to performing the main ML estimations
with \texttt{DeborahCore}, \texttt{Esther}, and \texttt{Miriam}.
It allows users to inspect correlations among observables, review the
HMC history of the input data, and estimate suitable block sizes for
block bootstrap analyses before the main computations are executed.
For example, Fig.~\ref{fig:corr-obs-1}, which appeared in the
preliminary discussion of correlations among observables, and
Fig.~\ref{fig:block-scan}, which was used in this work to determine
the provisional block size for the block-bootstrap resampling, were
both generated by the \texttt{Rahab} module.

\bibliographystyle{apsrev4-2}
\bibliography{ref}

\end{document}